\documentclass[12pt,preprint]{aastex}
\usepackage{amsmath}
\let\url\relax
\makeatletter
\renewcommand{\fps@figure}{tp}
\makeatother
\setkeys{Gin}{width=\linewidth,height=0.9\textheight,keepaspectratio=true}

\newcommand{\TA}{\tablenotemark{a}}
\newcommand\TB{\tablenotemark{b}}
\newcommand\TC{\tablenotemark{c}}

\newcommand{\elecd}{$n_{\rm e}$}
\newcommand{\elect}{$T_{\rm e}$}
\newcommand{\hb}{H$\beta$}
\newcommand{\ha}{H$\alpha$}

\newcommand{\foiii}{[\ion{O}{3}]}

\newcommand{\fsii}{[\ion{S}{2}]}

\newcommand{\fnii}{[\ion{N}{2}]}

\newcommand{\fcliii}{[\ion{Cl}{3}]}

\newcommand{\oiii}{\ion{O}{3}}

\newcommand{\nii}{\ion{N}{2}}

\newcommand{\oi}{\ion{O}{1}}
\newcommand{\oii}{\ion{O}{2}}

\newcommand{\cii}{\ion{C}{2}}

\newcommand{\sii}{\ion{S}{2}}

\newcommand{\feiii}{\ion{Fe}{3}}

\newcommand{\hi}{\ion{H}{1}}
\newcommand{\hii}{\ion{H}{2}}

\newcommand{\ts}{\emph{$t^{\rm 2}$}}

\shorttitle{The abundance discrepancy in the Orion nebula}
\shortauthors{Mesa-Delgado, Esteban \& Garc\'{\i}a-Rojas}

\received{}
\begin{document}

\title{Small scale behavior of the physical conditions\\
and the abundance discrepancy in the Orion nebula\footnotemark{}}

\author{Adal Mesa-Delgado, C\'esar Esteban, and Jorge Garc\'{\i}a-Rojas\footnote{Present address: Instituto de Astronom\'{\i}a, UNAM, Apdo. Postal 70-264,  
04510 M\'exico D.F., Mexico}}
\affil{Instituto de Astrof\'\i sica de Canarias, E-38200 La Laguna, Tenerife, 
Spain}
\email{amd@ll.iac.es, cel@ll.iac.es, jogarcia@ll.iac.es }

\begin{abstract}
We present results of long-slit spectroscopy in several positions of the Orion nebula. Our goal is to study the spatial distribution of a large number of nebular quantities, 
including line fluxes, physical conditions and ionic abundances at a spatial resolution of about 1$''$. In particular, we have compared the O$^{++}$ abundance 
determined from collisionally excited and recombination lines in 671 individual 1D spectra covering different morphological zones of the nebula. We find that protoplanetary 
disks (proplyds) show prominent spikes of {\elect}([{\nii}]) probably produced by collisional deexcitation due to the high electron densities found in these objects. 
Herbig-Haro objects show also relatively high {\elect}([{\nii}]) but probably produced by local heating due to shocks. We also find that the spatial distribution 
of pure recombination {\oii} and 
{\foiii} lines is fairly similar, in contrast to that observed in planetary nebulae. The abundance discrepancy factor (ADF) of O$^{++}$ remains rather constant along the slit positions, except in some particular small areas of the nebula where this quantity reaches somewhat higher values, in particular at the location of the most conspicuous Herbig-Haro objects: HH 202, HH 203, and HH 204. There is also an apparent slight increase of the ADF in the inner 40$''$ around $\theta^1$ Ori C. We find a negative radial gradient of {\elect}([{\oiii}]) and {\elect}([{\nii}]) in the nebula based on the projected distance from $\theta^1$ Ori C. We  explore the behavior of the ADF of O$^{++}$ with respect to other nebular quantities, finding that it seems to increase very slightly with the electron temperature. Finally, we estimate the value of the mean-square electron temperature fluctuation, the so-called {\ts} parameter. Our results indicate that 
the hypothetical thermal inhomogeneities --if they exist-- should be smaller than our spatial resolution element.  
\end{abstract}


\section{Introduction}

\footnotetext[1]{Based on observations made with the 4.2m William Herschel Telescope (WHT) operated 
on the island of La Palma by the Isaac Newton Group in the Spanish Observatorio del Roque de los Muchachos 
of the Instituto de Astrof\'\i sica de Canarias.}

\label{intro}

The analysis of the spectrum of {\hii} regions allows to determine the chemical composition of the ionized gas phase of the 
interstellar medium from the Solar Neighbourhood to the high-redshift Universe. Therefore, it stands as an essential tool 
for our knowledge of the chemical evolution of the Universe. 
In photoionized nebulae, the abundance of the elements heavier than He is usually determined from collisional excitation 
lines (hereinafter CELs), whose  intensity depends exponentially on the electron temperature, {\elect}, of the gas. It was 
about 20 years ago when the first determinations of C$^{++}$ abundance from the intensity of the weak 
recombination line (hereinafter RL) of {\cii} 4267 \AA\ were available for planetary nebulae (PNe). The comparison of the abundance obtained 
from {\cii} 4267 \AA\ and from the CELs of this ion in the ultraviolet (UV) showed a difference that could be as large as a order 
of magnitude in some objects \citep[e.g.][]{french83, rolastasinska94, mathisliu99}. \citet{peimbertetal93} were the first in determinig 
the O$^{++}$ abundance from the very weak RLs, obtaining the same qualitative result: the abundances 
obtained from RLs are higher than those determined making use of CELs. This observational fact is currently known as the ``abundance 
discrepancy" (hereinafter AD) problem. In the last years, our group has obtained a large dataset of intermediate and high 
resolution spectroscopy of Galactic and extragalactic {\hii} regions using medium and large aperture telescopes \citep{estebanetal02, 
estebanetal05, garciarojasetal04, garciarojasetal05, garciarojasetal06, garciarojasetal06b, lopezsanchezetal06}. 
The general result of 
these works is that the O$^{++}$/H$^+$ ratio calculated from RLs is between 0.10 and 0.35 dex higher than the value obtained from 
CELs in the same objects. The value of the AD that we usually find in {\hii} regions is rather similar for all objects and ions and 
is much lower than the most extreme values found in PNe. The results for {\hii} regions obtained by our group are fairly different that 
those found for PNe, and seem to 
be consistent with the predictions of the temperature fluctuations paradigm formulated by \citet{peimbert67}, 
as it is argued in \citet{garciarojas06} and \citet{garciarojasesteban07}. In the presence of 
temperature fluctuations (parametrized by the mean square of the spatial variations of temperature, the so-called $t^{\rm 2}$ parameter) 
the AD can be naturally explained because the different temperature dependence of the intensity of RLs and CELs. The existence and 
the origin of temperature fluctuations are still controversial problems and a challenge for our understanding of ionized nebulae. 
Recently, \citet{tsamispequignot05} and \citet{stasinskaetal07}  have proposed an hypothesis to the origin of the AD, which is based 
on the presence of cold high-metallicity clumps of supernova ejecta still not mixed with the ambient gas of the {\hii} regions. This 
cold gas would produce most of the emission of the RLs whereas the ambient gas of normal abundances would emit most of the intensity of 
CELs. 

Our group is interested in exploring on what variable or physical process the AD depends from different approaches. One of the most 
promising is based on the study of the behaviour of this magnitude at small spatial scales, something that has 
still not been explored in depth in nearby bright Galactic {\hii} regions. In this paper, we make use of deep intermediate-resolution 
long-slit spectroscopy of the Orion nebula to study the dependence of the AD 
with respect to different nebular parameters: electron temperature and density, local ionization state of the gas, presence of high 
velocity material, and its correlation with different morphological structures (proplyds, ionization fronts, globules, Herbig-Haro 
objects), in {\hii} regions.

The spatial distribution of the physical conditions in the Orion nebula has been investigated by several authors. 
\citet{baldwinetal91} obtained the density and temperature distribution in 21 and 14 points, respectively, along a 5$'$ line west 
of $\theta^1$ Ori C, finding a density gradient that decreases to the outskirts of the nebula and a constant {\elect}. \citet{walteretal92} 
determined electron densities and temperatures and chemical abundances for 22 regions of the Orion nebula. Using also data from the 
literature, these authors find radial gradients of the physical conditions, but with a positive slope in the case of the temperature 
determined from {\foiii} lines.  
\citet{poggeetal92} obtained Fabry-Perot images of the inner 6$'$ of the nebula covering several bright CELs and taken with an 
average seeing of about 1\farcs8. Those authors present a density map obtained from the ratio of the {\fsii} doublet confirming the 
presence of a density gradient that reaches its highest point immediately south-southwest of the Trapezium stars, and some localized 
density enhancements in the Orion bar and some Herbig-Haro objects. Very recently, \citet{sanchezetal07} have obtained an integral 
field spectroscopy mosaic of an area of 5$'\times 6'$ of the center of the Orion nebula, with a spatial resolution of 2\farcs7. 
The electron density map they obtain is consistent with that obtained by \citet{poggeetal92} but richer in substructures, some 
of them possibly associated to Herbig-Haro objects. \citet{sanchezetal07} also obtain a electron temperature map (derived from the line ratio 
of {\fnii} lines) that shows clear spatial variations, which rise near the Trapezium and drop to 
the outer zones of the nebula. However, an important drawback of the temperature map of S\'anchez et al. is that is based 
on non-flux calibrated spectra and possible effects due to variations in the dust extinction distribution cannot be disregarded. 
\citet{odelletal03} obtained a high spatial resolution map of the electron temperature --derived from the line ratio of {\foiii} 
lines-- of a 160$''\times 160''$ field centered at the southwest of the Trapezium. The data were obtained from narrow-band images 
taken with the WFPC of the  $HST$. Although they do not find a substantial radial gradient of 
{\elect} in the nebula, \citet{odelletal03} report the existence of small-scale temperature variations down to a few arcseconds compatible with 
the values of the temperature fluctuations parameter calculated from the AD determinations by \citet{estebanetal04}. \citet{rubinetal03} obtained   
$HST$/STIS long-slit spectroscopy at several slit positions on the Orion nebula analysing the electron temperature and density spatial 
profiles with resolution elements of 0\farcs5 $\times$ 0\farcs5. These last authors do not find large-scale gradients of the physical conditions 
along the slits but a relatively large point-to-point variation and some correlation of such variations with several small-scale structures. 

The spatial mapping of the AD factor has been performed in few ionized nebulae but largely for PNe. \citet{liuetal00}, 
\citet{garnettdinerstein01}, and \citet{krabbecopetti06} have found significant differences in the spatial profiles of the 
O$^{++}$/H$^+$ ratio derived making use of RLs and CELs suggesting the presence of chemical inhomogeneities or additional mechanisms 
for producing the {\oii} lines in these objects. \citet{tsamisetal03} have performed the only available study so far of the spatial 
distribution of the AD factor in an {\hii} region: 30 Doradus. However, considering the extragalactic nature of this object 
and the spatial sampling of 3\farcs5 used by those authors, their final spatial resolution is very low --about 1pc. In any case, 
\citet{tsamisetal03} find a rather constant AD factor along the zone covered with their observations, a quite different behavior than that 
observed in PNe. 

In \S\S~\ref{obsred} and~\ref{linesel} of this paper we describe the observations, the data reduction procedure and 
the aperture extraction and measurement of the emission lines. In \S~\ref{phiscondabund} we derive the physical conditions and the 
ionic abundances from both kinds of lines: CELs and RLs. In \S~\ref{spat_prof} we present and discuss 
the spatial profiles of the physical conditions, line fluxes, and the abundance discrepancy factor along the slit positions. 
In \S~\ref{rad_dist} we discuss the large-scale radial distribution of some nebular properties along the nebula. In \S~\ref{cor_ADF} we 
explore possible correlations between the AD and different nebular parameters. In \S~\ref{t2} we address and estimate the possible temperature 
fluctuations inside the nebula. Finally, in \S~\ref{conclu} we summarize our main conclusions. 

\section{Observations, Data Reduction, and Extraction of 1D Spectra}
\label{obsred}
Intermediate-resolution spectroscopy was obtained in 2002 December 27 with the ISIS spectrograph  
at the 4.2m William Herschel Telescope (WHT) in Observatorio del Roque de los Muchachos 
(La Palma, Spain). Two different CCDs were used at the blue and red arms of the spectrograph: 
an EEV CCD with a configuration 4096 $\times$ 2048 pixels with a size of 13.5 $\mu$m per pixel in the blue arm and a 
Marconi CCD with 4700 $\times$ 2148 pixels with a pixel size of 13.5 $\mu$m in the red arm. The dichroic prism used to separate 
the blue and red beams was set at 5400 \AA. The slit was 3\farcm7 long and 1\farcs03 wide. Two gratings 
were used, the R1200B in the blue arm and the R316R in the red arm. These gratings give reciprocal 
dispersions of 17 and 62 \AA\ mm$^{-1}$, and effective spectral resolutions of 0.86 and 3.81 \AA\ 
for the blue and red arms, respectively. The blue spectra cover from $\lambda\lambda$4198 to 5048 \AA\ and the red 
ones from $\lambda\lambda$5370 to 8690 \AA. The spatial scale is 0\farcs20 pixel$^{-1}$ in both arms. The average 
seeing during the observations was $\sim$1\farcs2.

\begin{figure*}
  \includegraphics[]{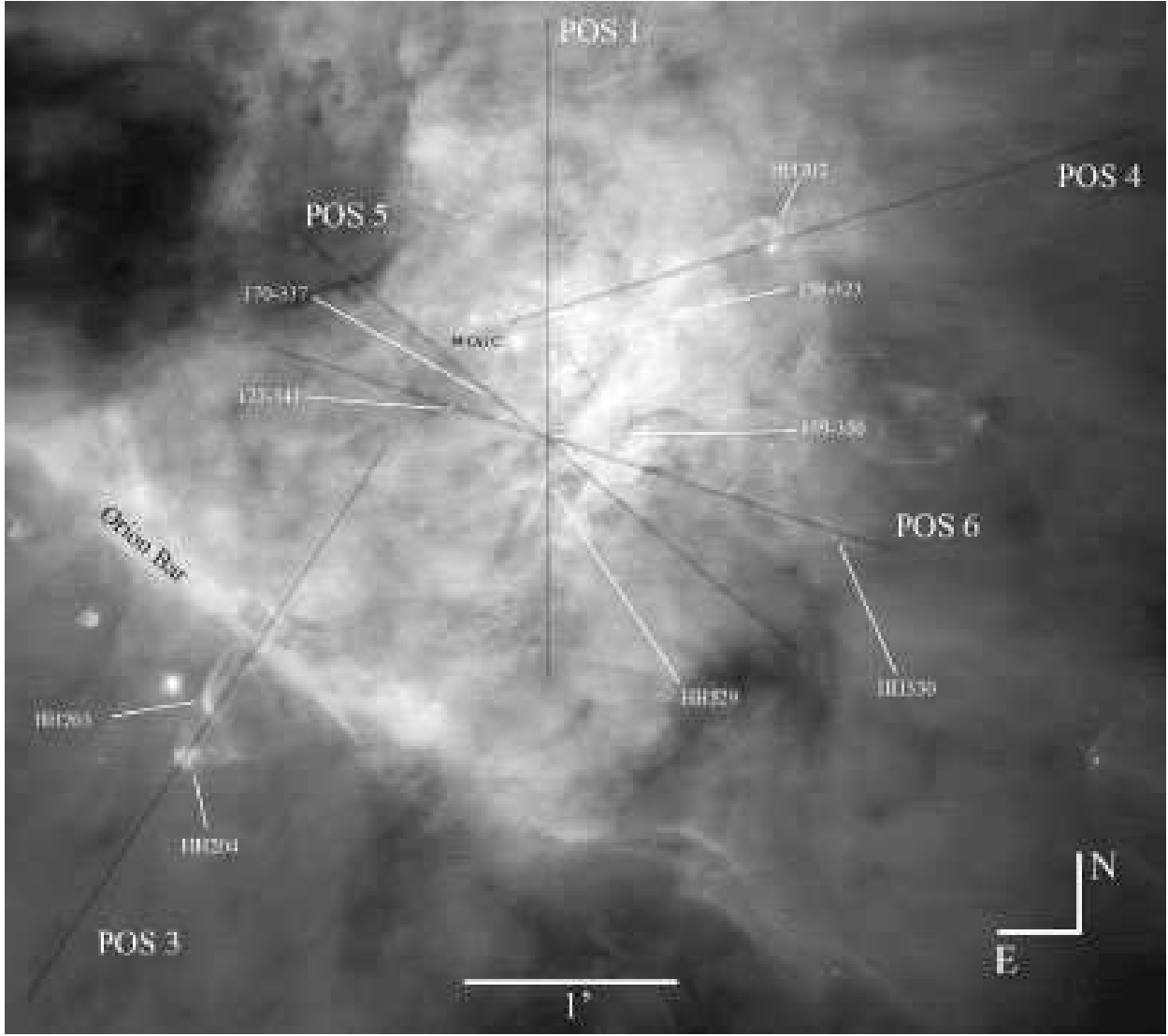}
  \caption{Our slit positions over a mosaic of a combination of WFPC2 images of the Orion nebula taken with different filters \citep{odellwong96}.}
  \label{slits}
\end{figure*}

We observed 5 slit positions covering different zones of the nebula and different position angles (see Figure~\ref{slits}). 
These positions were chosen in order to cover different morphological structures as proplyds (158-323, 158-326, 159-350, 170-337, and 177-341), 
Herbig-Haro objects 
(HH~202, HH~203, HH~204, HH~529, and HH~530) and the Orion bar. Due to the high surface brightness 
of the nebula, a large number of individual short exposures were taken in each slit position and spectral range in order to achieve a 
good signal-to-noise ratio in the faint {\cii} and {\oii} RLs and to avoid saturation of the brightest emission lines. The journal 
of observations can be found in Table~\ref{observations}. Note that the 5 slit positions are numbered as 1, 3, 4, 5, and 6, actually there 
was not a slit position number 2.  

The spectra were wavelength calibrated with a CuNe+CuAr lamp. The correction for atmospheric extinction was performed using the 
average curve for continuous atmospheric extinction at Roque de los Muchachos Observatory. The absolute flux calibration 
was achieved by observations of the standard stars Feige~15, Feige~110, H600, and Hz~44. All the CCD frames were reduced using 
the standard $IRAF$\footnote{IRAF is distributed by National Optical Astronomical Observatories, operated by the Associated 
Universities for Research in Astronomy,Inc., under contract to the National Science Foundation} TWODSPEC reduction package to perform 
bias correction, flat-fielding, cosmic-ray rejection, wavelength and flux calibration, and sky subtraction. 

\begin{deluxetable}{ccccccc}
\tabletypesize{\scriptsize}
\tablecaption{Journal of observations
\label{observations}}
\tablewidth{0pt}
\tablehead{
 & & & & & 
\colhead{Spectral}  & 
\colhead{Exposure} \\ 
\colhead{Slit}  & & &  
\colhead{P.A.}  & 
\colhead{$\Delta\lambda$} &
\colhead{Resolution} &
\colhead{Time} \\
\colhead{Position}  &   
\colhead{R.A.{\TA}}  & 
\colhead{Decl.{\TA}} &
\colhead{(deg)} &
\colhead{(\AA)} &
\colhead{(\AA\ pix$^{-1}$)} &
\colhead{(s)}}  
\startdata 
1 & 05 35 15.0 & $-$05 23 04 & 0 & 4198--5048 &  0.23 & 30$\times$60 \\
 & & & & 5370--8690 &  0.84 & 40$\times$30 \\	
3 & 05 35 22.4 & $-$05 25 09 & 147 & 4198--5048 &  0.23 & 15$\times$150 \\
 & & & & 5370--8690 &  0.84 & 25$\times$60 \\
4 & 05 35 12.4 & $-$05 22 44 & 107 & 4198--5048 &  0.23 & 31$\times$60 \\
 & & & & 5370--8690 &  0.84 & 40$\times$30 \\
5 & 05 35 15.8 & $-$05 23 33 & 50 & 4198--5048 &  0.23 & 20$\times$100 \\
 & & & & 5370--8690 &  0.84 & 39$\times$30 \\
6 & 05 35 15.3 & $-$05 23 38 & 72 & 4198--5048 &  0.23 & 20$\times$100 \\
 & & & & 5370--8690 &  0.84 & 41$\times$30 \\
\enddata
\tablenotetext{a}{Coordinates of the slit center}
\end{deluxetable}

The extraction of 1D spectra was done automatically through an $IRAF$ script, using the $apall$ task. First, we traced the apertures interactivelly 
by selecting the brightest object (star, proplyd or Herbig-Haro object) in each 2D spectrum. In all cases 
we adjusted a third-order spline function and obtained a typical $rms$ of the fit between 0.05 and 0.1 pixels. 
In the following step, we defined the apertures to extract for each 2D spectra by using the coefficients 
obtained in the aperture tracing. For all the slit positions, we extracted apertures of 6 pixels size in the spatial direction, 
that corresponds to an angular scale of 1\farcs2 --the average seeing during the night. Therefore, each aperture covers a size of 
1\farcs2 $\times$ 1\farcs03 in the spatial and spectral directions respectively. At the distance of the Orion nebula \citep[450 pc,][]{odell01},  
1$''$ corresponds to a linear size of 0.0022 pc, 6.8 $\times$ 10$^{-15}$ cm, or 450 A.U. The slit center in the red arm is some pixels displaced with 
respect to the slit center in the blue arm; this effect has 
been corrected in the extraction procedure, ensuring the same spatial covering in the blue and red ranges. We have discarded the apertures located near the 
edges of the CCD, resulting in a final number of 154 apertures extracted --individual 1D spectra-- for 
each slit position except numbers 1 and 4. In these two cases a star --$\theta^1$ Ori A-- fell into the slit, and we discarded the apertures contaminated by stellar emission 
(11 apertures in 
position 1 and 12 in position 4). In addition, the last 17 apertures at the northwest edge of position 4 were also discarded because  
the temperature sensitive [{\oiii}] 4363 \AA\ line was not detected due to the faintness of the spectra. The final total number of apertures 
extracted was 730.      
 
For each slit position, we extracted additional 1D spectra collapsing the whole slit --the whole extension of the 154 individual apertures-- 
but excluding different particular zones: a) only those apertures contaminated by stellar emission --these will be designated as 
``whole slit" spectra, or b) the same zones as in a) but also those apertures covering proplyds, Herbig-Haro objects, or  
having a very low surface brightness --these spectra are designated as ``background gas". In Table~\ref{ext_data},  
for each slit position, we summarize the number of apertures excluded due to stellar emission contamination or non detection of 
the [{\oiii}] 4363 \AA\ line (upper row), the total number of 
usable apertures (middle row), and the apertures that we consider representative of the ``background gas" (lower row).

\begin{deluxetable}{cccccc} 
\tabletypesize{\footnotesize}
\tablecaption{Number of apertures extracted
\label{ext_data}}
\tablewidth{0pt}
\tablehead{
 & 
\colhead{Pos 1}  & 
\colhead{Pos 3}  & 
\colhead{Pos 4}  & 
\colhead{Pos 5}  & 
\colhead{Pos 6}}  
\startdata 
No. of excluded apertures\TA	     & 11	& \nodata       & 29       & \nodata	& \nodata   \\
No. of extracted apertures\TB	     & 143	& 154   	& 125      & 154	& 154       \\
``Background gas" zones\TC  & 1-60,72-84,90-143 & 85-154 & 25-50,75-85 & 1-57,62-74,79,154 & 1-40,45-63,70-154
\enddata
\tablenotetext{a}{Contaminated by stellar emission or non detection of the [{\oiii}] 4363 \AA\ line.}
\tablenotetext{b}{Summed for producing the ``whole slit" spectra.}
\tablenotetext{c}{Summed for producing the ``background gas"spectra.}
\end{deluxetable}

\section{Emission Line Selection, Flux Measurements, and Reddening Correction}  
\label{linesel}
The emission lines considered in our analysis are indicated in Table~\ref{sec_lin}. These lines were selected according 
to the following criteria: 

\begin{itemize}
\item {\hi} lines --{\ha}, {\hb} and H$\gamma$--, which are used to compute the reddening correction and to re-scale the line intensity ratios 
of the red spectral range with respect to the blue one.

\item Ratios of CELs of various species, which are used to compute physical conditions --such as the auroral lines {\foiii} 4363 \AA\ or 
{\fnii} 5754 \AA\ to derive {\elect} or {\fsii} lines to derive {\elecd}-- and ionic abundances. 

\item Faint recombination lines of {\cii} and {\oii}, which are used to derive the C$^{++}$ and O$^{++}$ abundances and to compute the abundance 
discrepancy factor for O$^{++}$ (via a comparison with the O$^{++}$ abundances from CELs).

\item Some lines which are blended with other lines of interest.
\end{itemize}

\begin{deluxetable}{cccccc} 
\tabletypesize{\footnotesize}
\tablecaption{Selected lines
\label{sec_lin}}
\tablewidth{0pt}
\tablehead{
\colhead{$\lambda$ (\AA)}  & 
\colhead{Ion} & 
\colhead{Mult.}  & 
\colhead{$\lambda$ (\AA)}  & 
\colhead{Ion} & 
\colhead{Mult.}}
\startdata 
4267.15		& \ion{C}{2}	& 6       & 5517.71         & [\ion{Cl}{3}]	& 1F   \\
4340.47		& \ion{H}{1}	& H$\gamma$& 5537.88      & [\ion{Cl}{3}]	& 1F   \\
4363.21		& [\ion{O}{3}]	& 2F      & 5754.64       & [\ion{N}{2}]	& 3F   \\
4638.86		& \ion{O}{2}	& 1       & 6300.30	  & [\ion{O}{1}]	& 1F   \\
4640.64		& \ion{N}{3}	& 2       & 6312.10	  & [\ion{S}{3}]	& 3F   \\
4641.81		& \ion{O}{2}	& 1       & 6363.78	  & [\ion{O}{1}]	& 1F   \\
4643.06		& \ion{N}{2}	& 5       & 6548.03	  & [\ion{N}{2}]	& 1F   \\
4649.13		& \ion{O}{2}	& 1       & 6562.82	  & \ion{H}{1}  	& H$\alpha$   \\
4650.84		& \ion{O}{2}	& 1       & 6583.41	  & [\ion{N}{2}]	& 1F   \\
4661.63		& \ion{O}{2}	& 1       & 6716.47	  & [\ion{S}{2}]	& 2F   \\
4711.37		& [\ion{Ar}{4}]	& 1F      & 6730.85	  & [\ion{S}{2}]	& 2F   \\
4861.33		& \ion{H}{1}	& H$\beta$& 7135.78       & [\ion{Ar}{3}]	& 1F   \\
4881.00         & [\ion{Fe}{3}] & 2F      & 7319.19	  & [\ion{O}{2}]	& 2F   \\
4958.91		& [\ion{O}{3}]	& 1F      & 7330.20	  & [\ion{O}{2}]	& 2F   \\
5006.94		& [\ion{O}{3}]	& 1F      & 7751.10	  & [\ion{Ar}{3}]	& 2F   \\
\enddata
\end{deluxetable}

Line fluxes were measured applying a single or a multiple --in the case of line blending-- Gaussian profile fit 
procedure except in some apertures of positions 3 and 4 where the complex velocity field of the Herbig-Haro objects affects the 
line profiles and the Gaussian fit was not feasible. In these cases, the line intensities were measured by integrating all the flux 
included in the line profile between two given limits and over a local continuum estimated by eye. All the measurements were 
made with the SPLOT routine of the $IRAF$ package and using our own scripts to automatize the process.

To accurately compute the line fluxes we need to define the adjacent continuum of each line. This was done also 
by using the SPLOT routine. For each selected line, we define two small spectral zones --one at each side of the line-- and located 
as close as possible of the line and free of any spectral feature. SPLOT fits the continuum between both zones and obtains the flux 
of the line. This automatic procedure was tested in randomly selected lines where we manually measured the flux using a local continuum 
estimated by eye. In general, there is a good agreement between both kinds of measurements within the adopted uncertainties. 
The observational errors associated with the line flux measurements was determined following \citet{castellanosetal02} from the expression 
$\sigma_{\rm l} = \sigma_{\rm c} N^{1/2}[1+{\rm EW}/(N\Delta)]^{1/2}$, where $\sigma_{\rm l}$ is the error in the line flux, $\sigma_{\rm c}$ 
represents the standard deviation of the continuum close to the measured emission line, $N$ is the number of pixels used in the measurement of 
the line flux, EW is the line equivalent width, and $\Delta$ is the wavelength dispersion in \AA\ pixel$^{-1}$. The final uncertainty of the 
line intensity ratios are estimated to be tipically: 
about 1\% if the ratio $F$($\lambda$)/$F$({\hb})$\geq$0.1, 
about 2\% if 0.01$\leq$ $F$($\lambda$)/$F$({\hb}) $\leq$0.1, 
about 10\% if 0.005$\leq$ $F$($\lambda$)/$F$({\hb})$\leq$0.01, 
about 25\% if 0.001$\leq$ $F$($\lambda$)/$F$({\hb})$\leq$0.005 
and about 40\% if 0.0001$\leq$ $F$($\lambda$)/$F$({\hb})$\leq$ 0.001. 
We do not consider lines weaker than 0.0001 $\times$ $F$({\hb}) (see below). 

All the line intensities of a given aperture have been normalized to a particular {\hi} recombination 
line present in each wavelength interval. For the blue spectra, the reference line was 
{\hb}, and for the red one, the reference was {\ha}. To produce a final homogeneous set of line intensity ratios, 
all of them were re-scaled to {\hb}. The re-scaling factor used in the red spectra was the theoretical 
{\ha}/{\hb} ratio for the physical conditions of {\elect} = 10000 K and {\elecd} = 1000 cm$^{-3}$.
 
As it can be expected, weak lines were not detected in all the apertures. To avoid bad weak line measurements, we imposed three criteria 
to discriminate between real features and noise. The criteria are the following:

\begin{itemize}
\item Line intensity peak over 2.5 times the sigma of the continuum (2.5${\bf \sigma}$): The more usual criterion of 3$\sigma$ was 
not adopted because several important lines clearly detected were discarded using that constraint. 

\item FWHM($\lambda$) $>$ 1.5 $\times$ FWHM(H I) or FWHM($\lambda$) $<$  FWHM(H I)/1.5: These inequalities were used to discriminate between true 
emission lines and spurious features.

\item $I$($\lambda$) $<$ 0.0001$\times$$I$(H$\beta$). This intensity was near the detection limit of our observations.
\end{itemize}

\begin{figure}
  \includegraphics[angle=90,scale=0.8]{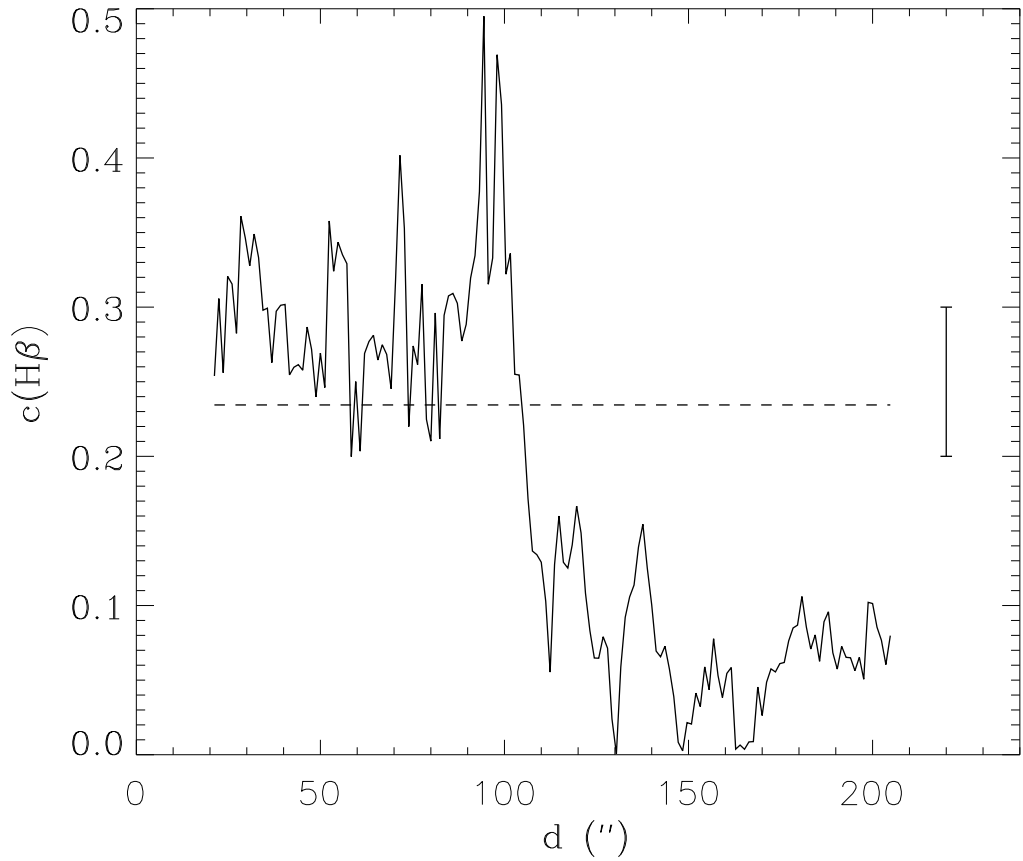} 
  \caption{Spatial distribution of the reddening coefficient, $c$({\hb}), along slit position 6. Positional measurement along the slit goes
 from northeast to southwest (see Figure~\ref{slits}). The dashed horizontal line represents the value of $c$({\hb}) obtained 
from the integrated spectrum along the whole slit. The typical error bar is included.}
  \label{chb_pos6}
\end{figure}

The reddening coefficient, $c$({\hb}), was obtained by fitting the observed H$\gamma$/{\hb} ratio to the theoretical one predicted by \citet{storeyhummer95} 
for the nebular conditions determined in the slit position observed by 
\citet{estebanetal04}. Following 
\citet{estebanetal98} we have used the reddening function, $f$($\lambda$), normalized at {\hb} derived by 
\citet{costeropeimbert70} for the Orion nebula. In Figure~\ref{chb_pos6}, as an example, we show the spatial distribuition of $c$({\hb}) 
obtained for the apertures of the slit position 6. The zone of the largest reddening at the left half of Figure~\ref{chb_pos6} corresponds 
to the northeast portion of slit position 6. The reddening maps of \citet{odellyusef-zadeh00} and \citet{sanchezetal07} also show larger values 
at this zone, which is in the vicinity of the so-called Dark Bay. The typical uncertainty of $c$({\hb}) is estimated to be about 0.05.

\section{Physical Conditions and Chemical Abundances}
\label{phiscondabund} 
\subsection{Electron Temperatures and Densities}
\label{temden}
Nebular electron temperatures, {\elect}, and densities, {\elecd}, have been derived from the usual CEL ratios, using the {\sc IRAF} task
\emph{temden} of the package \emph{nebular} \citep{shawdufour95} with updated atomic data \citep[see][]{garciarojasetal05}. We have 
computed {\elecd} from the [{\sii}] 
6717/6731 line ratio and {\elect} from the nebular to auroral [{\oiii}] (4959+5007)/4363 
and [{\nii}] (6548+6584)/5754 line ratios. The spatial distributions of the physical 
conditions are presented and discussed in \S\ref{phys_cond}. Although we include the {\fcliii} 
doublet in our set of selected lines we do not use the lines in the analysis because the number of apertures with good determinations of the 
density sensitive line ratio is rather low.  

The methodology for the determination of the physical conditions was the following: an initial {\elect} of 10000 K
was assumed in order to derive a first approximation to {\elecd}([{\sii}]); then, the obtained {\elecd} was used to compute {\elect}([{\oiii}]) and
{\elect}([{\nii}]), and finally we iterated until convergence to compute the finally adopted values of {\elecd}
and {\elect}. Uncertainties in the physical conditions were computed by propagating the errors in the analytical
expression of {\elecd} computed by \citet{castanedaetal92} and that of {\elect} given by \citet{osterbrockferland06} (their equations 5.4 an 5.5). 
Although the expression derived by \citet{castanedaetal92} is only valid to a limited range of densities lower than 10$^4$ cm$^{-3}$, it seems adequate 
for simply estimating the error propagation due to uncertainties in the computed temperatures and line ratios.

\subsection{Ionic Abundances from CELs and RLs}
\label{abund} 
Ionic abundances of N$^+$, O$^+$, O$^{++}$, S$^+$, S$^{++}$, and Ar$^{++}$ have been derived from CELs making use of the {\sc IRAF} task
\emph{ionic} of the package \emph{nebular}. We have assumed a two-zone scheme and $t^{\rm 2}$ = 0, adopting {\elect}([{\nii}]) for ions with 
low ionization potential (N$^+$, O$^+$ and S$^+$) and {\elect}([{\oiii}]) for ions with high ionization potential (O$^{++}$, S$^{++}$ and 
Ar$^{++}$). The errors in the ionic abundances are the quadratic sum of the independent contributions of {\elecd}, {\elect}, and line flux 
uncertainties.

\begin{figure}
\centering
  \includegraphics[angle=0]{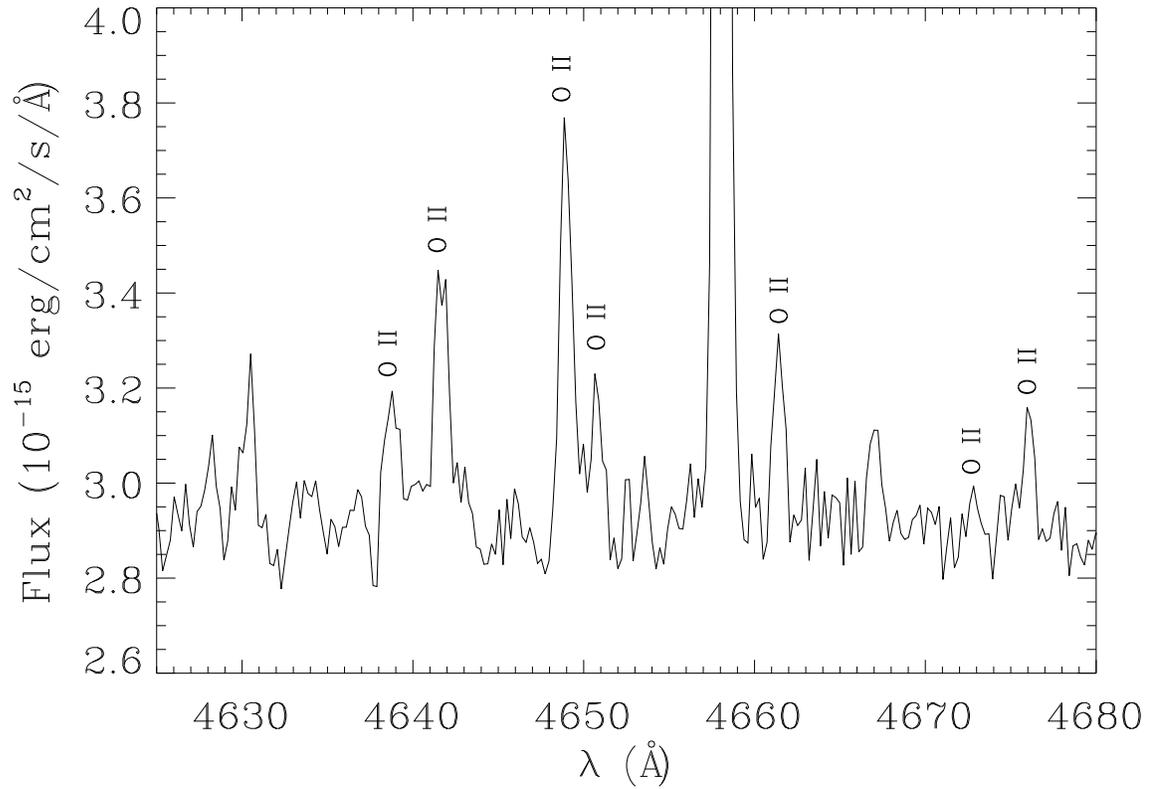} 
  \caption{Portion of the blue spectrum of an aperture extracted from slit position 6 and extending from the positional measurements 
151.4 to 152.6 arcseconds (see Figure~\ref{sp_pos6}). This is a representative example of the average quality of our 1D spectra.}
  \label{oii}
\end{figure}

On the other hand, the high signal-to-noise of the spectra has permitted us to detect and measure pure RLs of {\oii} and {\cii} 
in most of the apertures (see Figure~\ref{oii}). These lines have the advantage that their relative intensity with respect to \ion{H}{1} lines 
depend weakly on {\elect} and {\elecd}, avoiding the problem of the presence of temperature variations along the line of sight, that can actually affect 
the abundance determinations from CELs, which emissivities depend strongly on the {\elect} of the nebula.

Let $I$($\lambda$) be the intensity of a RL of a element X, $i$ times ionized, at wavelength $\lambda$; then the abundance of 
the ionization state $i+1$ of element X is given by
 \begin{eqnarray}
  \frac{N({\rm X}^{i+1})}{N({\rm H}^+)} = \frac{\lambda(\AA)}{4861} \frac{\alpha_{eff}({\rm H}\beta)}{\alpha_{eff}(\lambda)} \frac{I(\lambda)}{I({\rm H}\beta)},
 \end{eqnarray}
where $\alpha_{eff}(\lambda)$ and $\alpha_{eff}({\rm H}\beta)$ are the effective recombination coefficients for the line and H$\beta$, 
respectively. The $\alpha_{eff}({\rm H}\beta)/\alpha_{eff}(\lambda)$ ratio is almost independent of the adopted temperatures and densities.

Following \citet{estebanetal98} we have considered the abundances obtained from the intensity of each individual line of 
multiplet 1 of {\oii} and the abundances from the estimated total intensity of the multiplet. This last quantity is obtained 
by multiplying the sum of the intensities of the individual lines observed by the multiplet correction factor, defined as:   
 \begin{eqnarray}
  m_{cf} = \frac{\sum_{all} s_{ij} }{\sum_{obs} s_{ij}},
 \end{eqnarray}
where $s_{ij}$ is the theorical line strength, which are constructed assuming that they are proportional to the population of their parent levels 
assuming LTE computations predictions. The upper sum runs over $all$ the lines of the multiplet and the lower sum runs over the $observed$ lines of the multiplet. 

The O$^{++}$ and C$^{++}$ abundances from RLs have been calculated using the representative {\elect} of
these ions --{\elect}([{\oiii}])-- and the effective recombination coefficients available in the literature (\citealt{storey94} for 
{\oii} assuming LS coupling, and \citealt*{daveyetal00} for {\cii}). The 
O$^{++}$ abundance has only been computed when at least three lines of multiplet 1 were measured in a given 1D spectrum.  The final 
number of apertures with determinations of the O$^{++}$/H$^+$ ratio obtained from RLs was 671, a 92\% of the total number of available 1D spectra.
NLTE corrections 
are not taken into account for deriving the O$^{++}$ abundances \citep[see][]{tsamisetal03,ruizetal03,peimbertetal05} considering that we use several lines of the multiplet 
and the high densities --between 4000 and 6000 cm$^{-3}$ \citep[e.g.][]{estebanetal98, estebanetal04}-- of the Orion nebula.

\section{Spatial Profiles along the Slit Positions}
\label{spat_prof} 

\subsection{Physical Conditions}
\label{phys_cond} 
The first step in the analysis of our results was the obtaining of spatial profiles of several nebular parameters along the slit positions. The selected 
parameters were: c(H$\beta$), {\elecd}, {\elect}([{\nii}]), {\elect}([{\oiii}]), the intensity of several selected lines (H$\beta$, {\cii} 4267 \AA,
{\oii} 4649 \AA, [{\oiii}] 4959 \AA, [{\feiii}] 4881 \AA, [{\nii}] 5755 and 6584 \AA,
[{\oi}] 6300 \AA, and [{\sii}] 6717 + 6731 \AA), the O$^{++}$/H$^+$ ratio obtained from CELs and RLs, and the C$^{++}$/H$^+$ ratio obtained from RLs. 
In figures 4 to 8 we show some selected spatial profiles of slit positions 3, 6, 1, 4, and 5, respectively. Slit positions 3 
and 6 are the most interesting ones and we will focus our discussion on their main features.  
Slit position 3 crosses the Orion bar and the Herbig-Haro (H-H) objects HH 203 and HH 204 and slit position 6 passes through the brightest part of the nebula at the 
southwest of the Trapezium cluster, two proplyds: 159-350 and 177-341, and HH 530. 

The spatial profiles of {\elecd} show a large range of variation across the slits, with local maxima associated with the position of proplyds, H-H objects, the Orion bar, 
and the bright zone at the southwest of the Trapezium (see figures 4a to 8a). The highest densities are found at the proplyd 159-350 that has been observed in slit 
positions 5 and 6 
(see figures 8a and 5a, respectively). 
This object shows {\elecd} of the order of 6 $\times$ 10$^4$ and 2 $\times$ 10$^4$ cm$^{-3}$ in positions 5 and 6, respectively; whereas the proplyd 158-326 
--near $\theta^1$ Ori C-- 
shows values somewhat larger than 4 $\times$ 10$^4$ cm$^{-3}$ (See Figure 6a). The densities at the brightest zone of the nebula --the southwest of the Trapezium--  
are about 2.5$\times$10$^4$ cm$^{-3}$ (see Figure 5a). It is obvious that so high {\elecd} determinations based on the [{\sii}] doublet are not totally confident 
because they are at the 
high-density limit of this indicator. The H-H objects are also associated with local peaks of {\elecd}, but not as high as in the proplyds, in fact HH 202, HH 203, 
and HH 204 show maxima between 6000 and 1 $\times$ 10$^4$ cm$^{-3}$. In Figure 4a, the dashed line marks the {\elecd} obtained from the ``whole slit'' 
(integrated) spectrum --2460 cm$^{-3}$-- whereas the minimum and maximum values are about 700 and 8000 cm$^{-3}$, covering a range of one order of magnitude. 
In Figure 5a, we can see that the range of variation of {\elecd} is also dramatic along slit position 6. 
In this figure we also compare the values corresponding to the ``whole slit'' spectrum 
--5500 cm$^{-3}$-- and the ``background gas" --4700 cm$^{-3}$-- one, that corresponds to an integrated spectrum excluding the emission of the proplyds. 
As we can see, the measured density increases 
in 800 cm$^{-3}$ --about a 17\%-- when we include the emission of the proplyds in the integrated spectrum of the slit. 

\begin{figure*}
  \includegraphics[angle=90,scale=0.8]{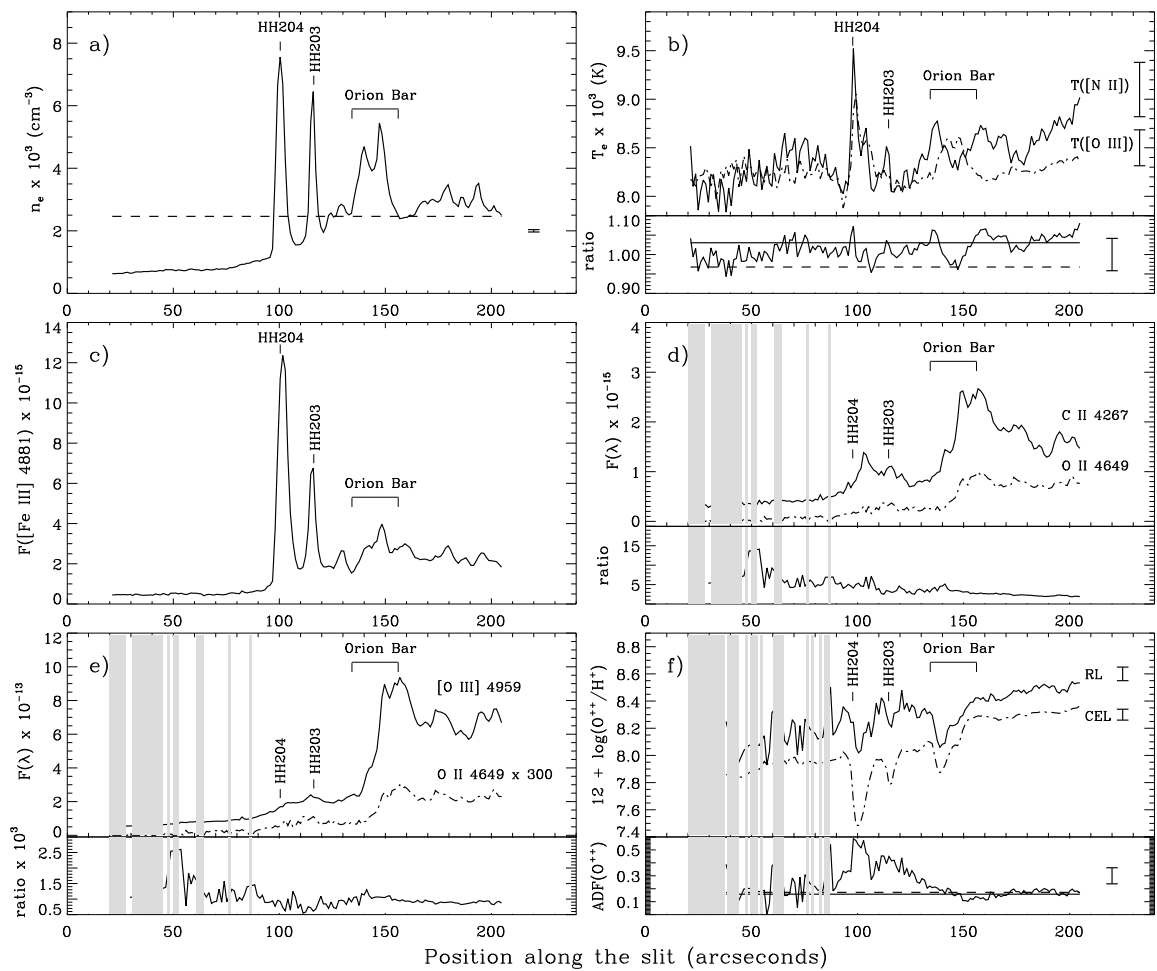} 
  \caption[spatial profiles pos3]{Spatial profiles of several nebular parameters along slit position 3. Positional measurement along the slit goes
 from southeast to northwest (see Figure~\ref{slits}). The position of the Orion bar and the Herbig-Haro objects HH 203 and HH 204 
are indicated. Typical error bars are included in some of the diagrams. 
(a) {\elecd}, the horizontal long-dashed line gives the average value of {\elecd} for the ``whole slit'' (integrated) spectrum. 
(b) upper pannel: {\elect}([{\oiii}]) (dashed-dotted line) and {\elect}([{\nii}]) (continuous line); lower pannel: 
{\elect}([{\nii}])/{\elect}([{\oiii}]) ratio, the horizontal continuous and long-dashed lines represent the value of that ratio for 
the ``background gas" and `whole slit'' spectra, respectively. (c) $F$([{\feiii}] 4881). (d) upper pannel: $F$({\cii} 4267) 
(continuous line) and $F$({\oii} 4649) (dashed-dotted line); lower pannel:  $F$({\cii} 4267)/$F$({\oii} 4649) ratio, the vertical grey lines indicate 
zones without a reliable measurement of the {\oii} 4649 line. (e) upper pannel: $F$({\foiii} 4959) (continuous line) and $F$({\oii} 4649) 
(dashed-dotted line), lower pannel: $F$({\foiii} 4959)/$F$({\oii} 4649) ratio, the vertical grey bands indicate zones without a reliable measurement of the 
{\oii} 4649 line. (f) upper pannel: 12 + log(O$^{++}$/H$^+$) determined from RLs (continuous 
line) and CELs (dashed-dotted line); lower pannel: ADF(O$^{++}$), the straight continuous and long-dashed lines represent the value of the ADF for 
the ``background gas" and `whole slit'' spectra, respectively, the vertical grey bands indicate zones without a reliable determination of the O$^{++}$/H$^+$ ratio 
from RLs.}
  \label{sp_pos3}
\end{figure*}

\begin{figure*}
  \includegraphics[angle=90,scale=0.8]{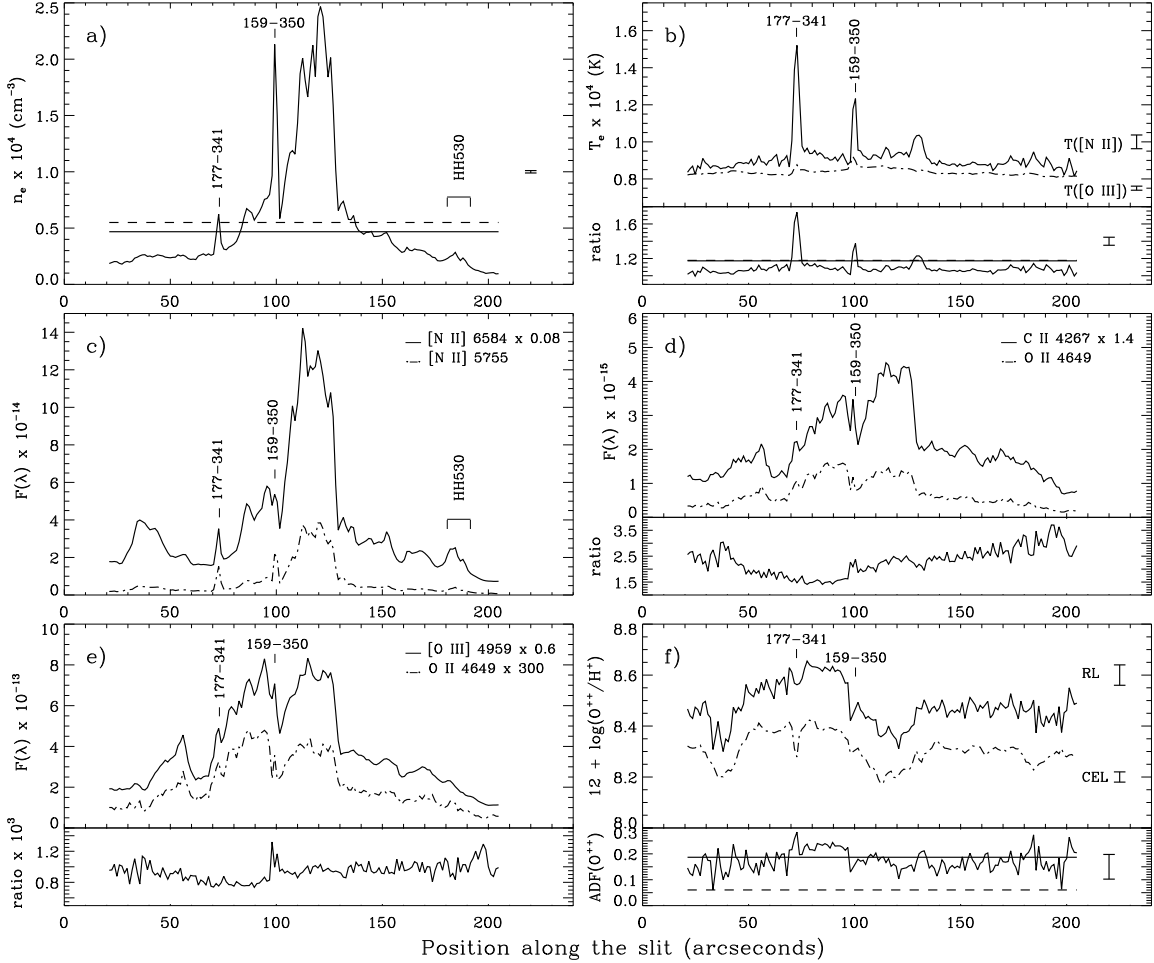} 
  \caption[spatial profiles pos6]{Spatial profiles of several nebular parameters along slit position 6. Positional measurement along the slit goes
 from northeast to southwest (see Figure~\ref{slits}). The position of the proplyds 177-341 and 159-350 and the Herbig-Haro object HH 530  
are indicated. Typical error bars are included in some of the diagrams.
(a) {\elecd}, the horizontal continuous and long-dashed lines represent the value of {\elecd} for 
the ``background gas" and `whole slit'' spectra, respectively. 
(b) upper pannel: {\elect}([{\oiii}]) (dashed-dotted line) and {\elect}([{\nii}]) (continuous line); lower pannel: 
{\elect}([{\nii}])/{\elect}([{\oiii}]) ratio, the horizontal continuous and long-dashed lines represent the value of that ratio for 
the ``background gas" and `whole slit'' spectra, respectively. (c) $F$([{\fnii}] 5755) (continuous line) and 
$F$([{\fnii}] 6584) (dashed-dotted line). (d) upper pannel: $F$({\cii} 4267) 
(continuous line) and $F$({\oii} 4649) (dashed-dotted line); lower pannel:  $F$({\cii} 4267)/$F$({\oii} 4649) ratio. 
(e) upper pannel: $F$({\foiii} 4959) (continuous line) and $F$({\oii} 4649) (dashed-dotted line); 
lower pannel: $F$({\foiii} 4959)/$F$({\oii} 4649) ratio. 
(f) upper pannel: 12 + log(O$^{++}$/H$^+$) determined from RLs (continuous 
line) and CELs (dashed-dotted line); lower pannel: ADF(O$^{++}$), the horizontal continuous and long-dashed lines represent the value of the ADF for 
the ``background gas" and ``whole slit'' spectra, respectively.}
  \label{sp_pos6}
\end{figure*}

\begin{figure*}
  \includegraphics[angle=90,scale=0.8]{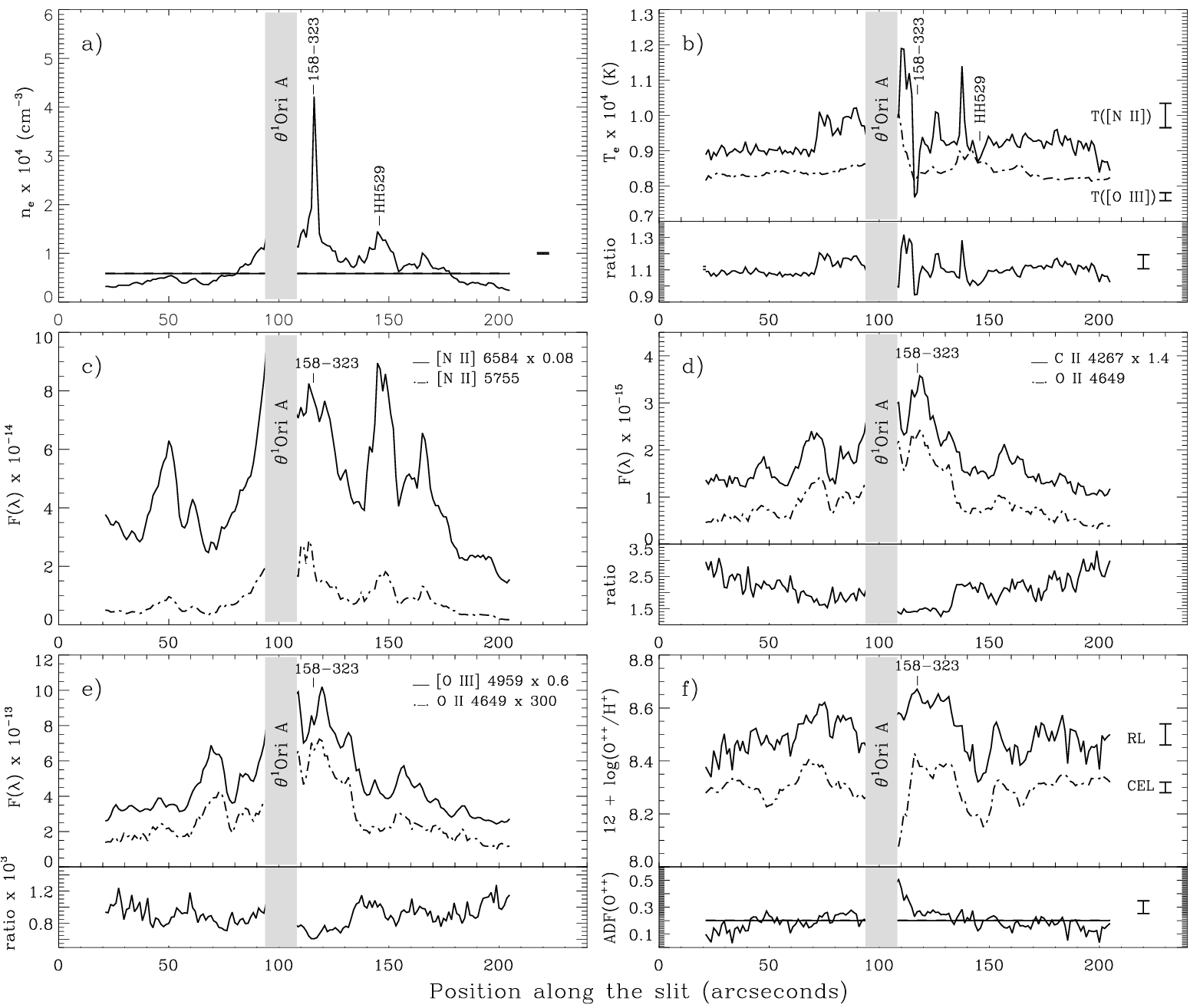} 
  \caption[spatial profiles pos1]{The same as Figure~\ref{sp_pos6} but for slit position 1. Positional measurement along the slit goes
 from north to south (see Figure~\ref{slits}).}
  \label{sp_pos1}
\end{figure*}

\begin{figure*}
  \includegraphics[angle=90,scale=0.8]{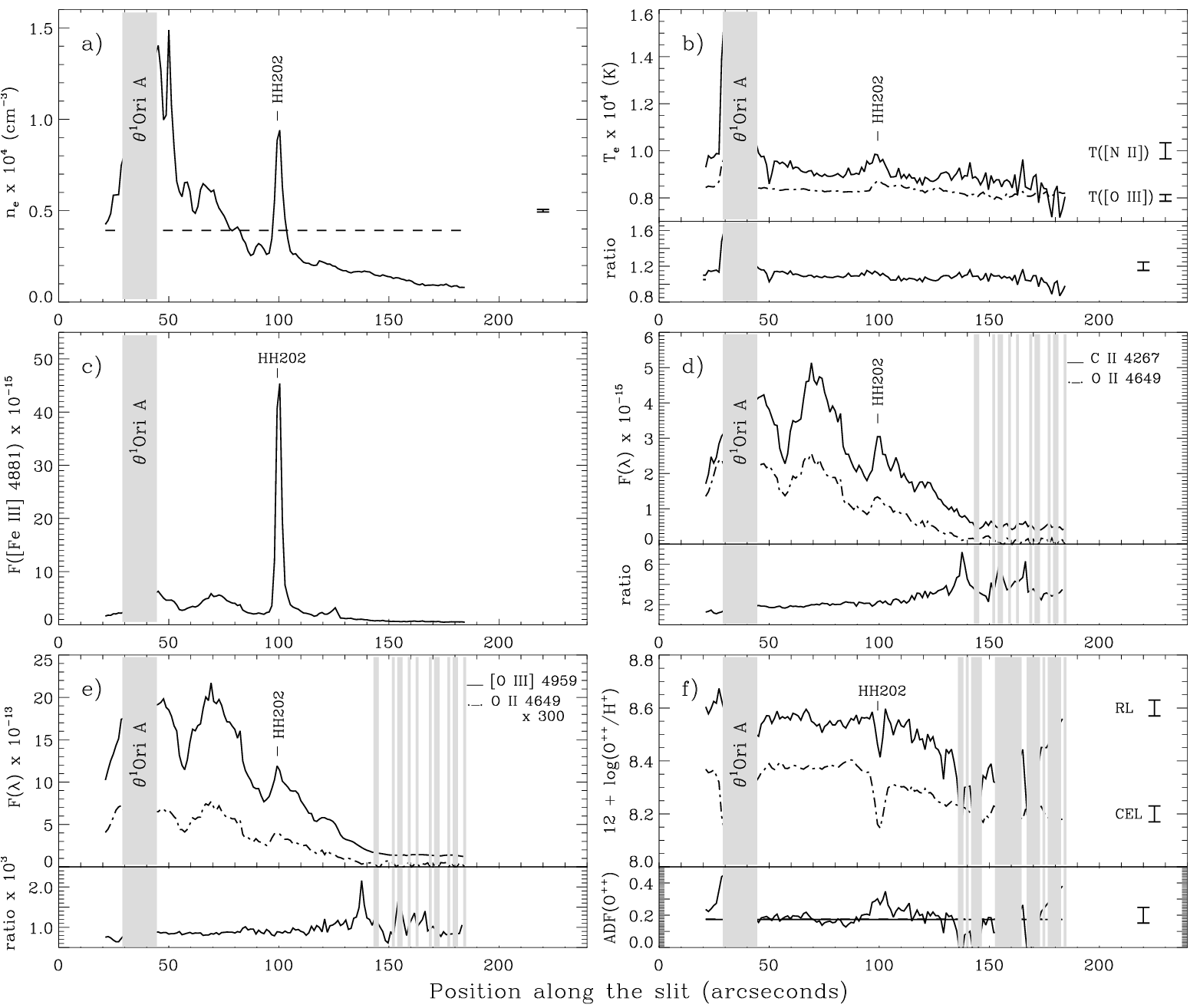} 
  \caption[spatial profiles pos4]{The same as Figure~\ref{sp_pos3} but for slit position 4. Positional measurement along the slit goes
 from southeast to northwest (see Figure~\ref{slits}).}
  \label{sp_pos4}
\end{figure*}

\begin{figure*}
  \includegraphics[angle=90,scale=0.8]{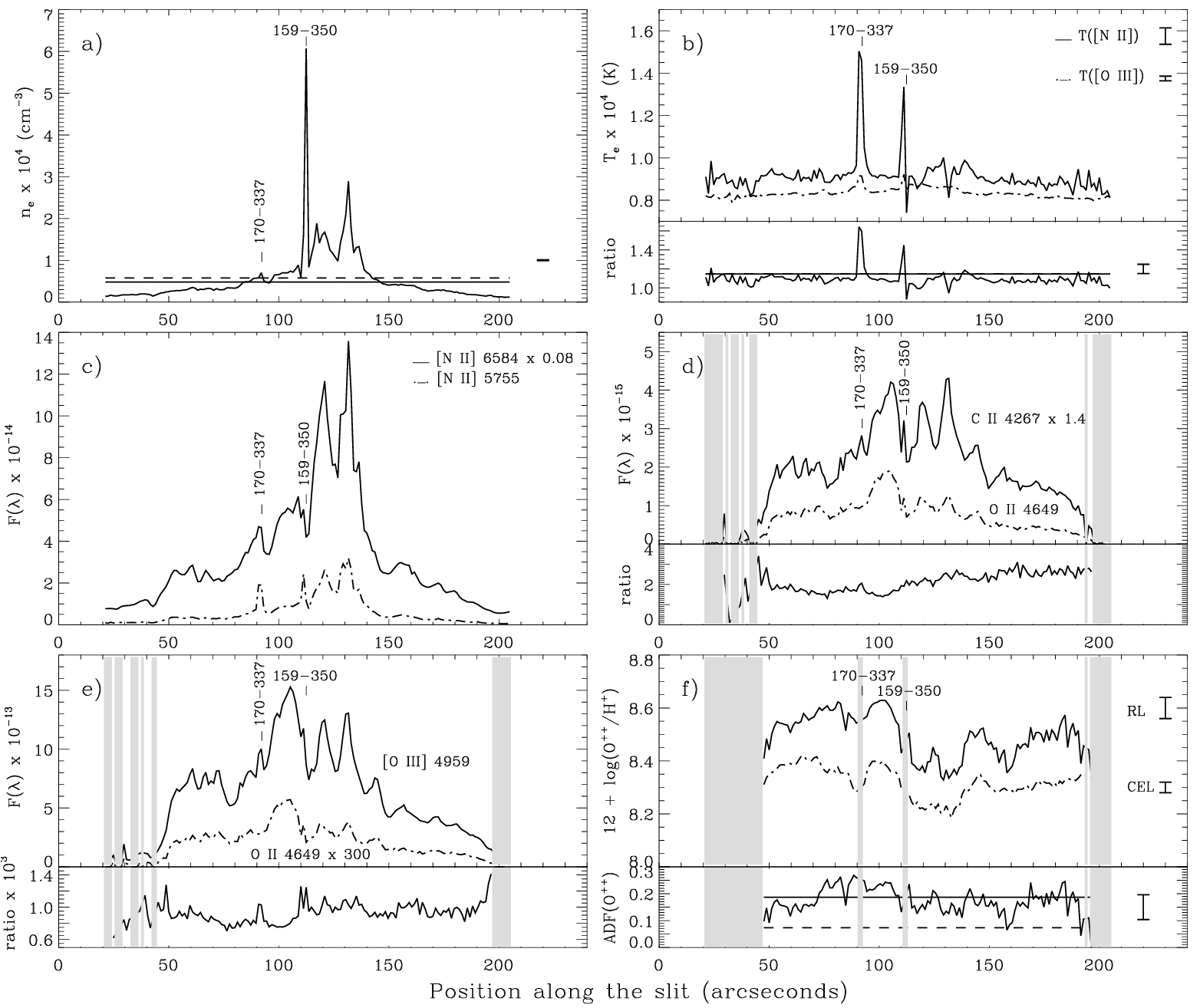} 
  \caption[spatial profiles pos5]{The same as Figure~\ref{sp_pos6} but for slit position 5. Positional measurement along the slit goes
 from northeast to southwest (see Figure~\ref{slits}).}
  \label{sp_pos5}
\end{figure*}

The spatial profiles of {\elect}([{\nii}]) and {\elect}([{\oiii}]) show very interesting features (see figures 4b to 8b). The proplyds observed in  
positions 1, 5, and 6 show quite prominent spikes of {\elect}([{\nii}]) and lesser or almost absent ones of {\elect}([{\oiii}]). In slit position 6 
(Figure 5b) the {\elect}([{\nii}]) increases locally about 70\% at proplyd 177-341, a similar spike shows 
170-337 in slit position 5 (Figure 8b). The {\elect}([{\nii}]) is also higher than the mean by 50\% and 40\% at the 
position of proplyd 159-250 in slit positions 5 and 6, respectively. In all the proplyds, the increase of {\elect}([{\oiii}]) 
is only of a few hundred K at most. In Figure 5b, we can also see a relatively broad  --about 5$''$ wide-- spike of {\elect}([{\nii}]) at 130$''$, 
where the temperature increases about 15\%. 
This feature is not related to any local structure reported by \citet{ballyetal00} but with a conspicuous dark globule that can be seen at the edge of the 
bright zone at the southwest of the Trapezium. There are also less important temperature spikes related to some H-H objects. 
In Figure 4b, we can see 
that {\elect}([{\nii}]) increases about 15\% at HH 204 and that the increase is slightly lower in the case of {\elect}([{\oiii}]). 
In contrast, HH 203 do not show that behavior and its temperatures 
are similar to those of the surrounding gas. In HH 202 --slit position 4-- the increase of {\elect}([{\nii}]) and {\elect}([{\oiii}]) are only about 
8\% and 5\%, respectively (Figure 7b). 
The other H-H objects we have observed --HH 529 and HH 530-- do not show temperature variations with respect to the surrounding gas (figures 6b and 5b). 
The different behavior of the 
temperatures in the H-H objects does not seem to be 
correlated with the velocity of their associated flows as reported by \citet{henneyetal07}. 
There is a last interesting feature regarding the temperature profiles 
that can be seen in Figure 4b. Although {\elect}([{\nii}]) is almost always a few hundred K larger than {\elect}([{\oiii}]) in all the slit positions 
\citep[this has been also observed in previous works, e.g.][]{baldwinetal91,rubinetal03}, the zone around the Orion bar --between 130$''$ and 150$''$ in Figure 4b-- 
shows a reversal of this relation just in the inner 10$''$ of the bar. In contrast, {\elect}([{\nii}]) shows a local increase just outside the bar. 
This local increment of {\elect}([{\oiii}]) we see in this particular zone could be related with the highly ionized jet --specially bright in 
{\foiii}-- that leads to HH 203 and HH204 \citep[see][]{doietal04}. In fact, the position of this zone coincides with a knot of high {\foiii} emission 
that can be seen in figure 8 of \citet{doietal04} that is crossed by our slit (their box with the position-based identifier 209-446). 

\citet{rubinetal03} 
obtained $HST$/STIS spectroscopy of a slit position very similar to our position 3 --their slit 4-- and show its {\elect}([{\oiii}]) spatial profile in their 
figure 3a. 
If we compare that figure with our Figure 4b we can see that the point-to-point dispersion of the temperature is substantially lower in our data. 
In fact, whereas {\elect}([{\oiii}]) 
varies from 6500 to 12000 K in the slit 4 of \citet{rubinetal03} the variations are only from 8000 to 9000 K in our slit position 3. 
This fact has two possible explanations: a) the presence of real temperature variations with typical spatial scale between 0\farcs5 (the spatial resolution of Rubin et al. 
data) and 1\farcs2 (our resolution) in the plane of the sky, or b) the temperature variations reported by \citet{rubinetal03} 
are spurious and produced by the much lower signal-to-noise ratio of their data. We think that the second explanation is perhaps the most likely one considering   
that the deepest exposures obtained by Rubin et al. for the spectral range containg {\foiii} 4363 \AA\ were about 1060 s and our exposures were between 1800 and 2200 s 
long. Moreover, an important factor should be added due to the different apertures of the telescopes used in our and Rubin et al.'s datasets --4.2m and 2.4m, respectively-- 
as well as the smallest element of spatial resolution used in Rubin et al.'s observations.  
 
It is clear that the behavior of {\elect}([{\nii}]) and {\elect}([{\oiii}]) at the positions of proplyds and H-H objects is different.
In figures 4b to 8b  
we also include the ratio of both temperatures showing clearly that the increase of {\elect}([{\nii}]) is higher than that of {\elect}([{\oiii}]) in proplyds. 
The presence of {\elect}([{\nii}]) enhancements in the proplyds of the Orion nebula was previously reported by \citet{rubinetal03} and they interpret this as the effect of 
collisional deexcitation on the nebular lines of [{\nii}] due to the high densities of these objects. We have explored that possibility comparing the spatial 
profile of the intensity of [{\nii}] 5755 and 6584 \AA\ lines. Both lines come from upper levels with very different critical densities 
--7.9$\times$10$^6$ cm$^{-3}$ in the case of [{\nii}] 5755 \AA\ line and 5.8$\times$10$^4$ cm$^{-3}$ in the case of [{\nii}] 6584 \AA\ line. 
In figures 5c and 6c, we can see that the two proplyds with the largest electron densities ({\elecd} $>$ 2-6$\times$10$^4$ cm$^{-3}$): 159-350 (also observed in Figure 8c) 
and 158-323 show a spike in the brightness of the [{\nii}] 5755 \AA\ line, whereas the [{\nii}] 6584 \AA\ line do not show a so clear increase in its brightness 
with respect to the emission of the surroundings. 
However, the proplyds showing the lowest electron densities ({\elecd} $<$ 1$\times$10$^4$ cm$^{-3}$) 177-341 and 170-337 
(figures 5c and 8c) show similar localized enhancements of the intensity of both {\fnii} lines, indicating that collisional deexcitation seems to be not so 
important in these 
two proplyds. 

\begin{figure}
  \includegraphics[angle=0]{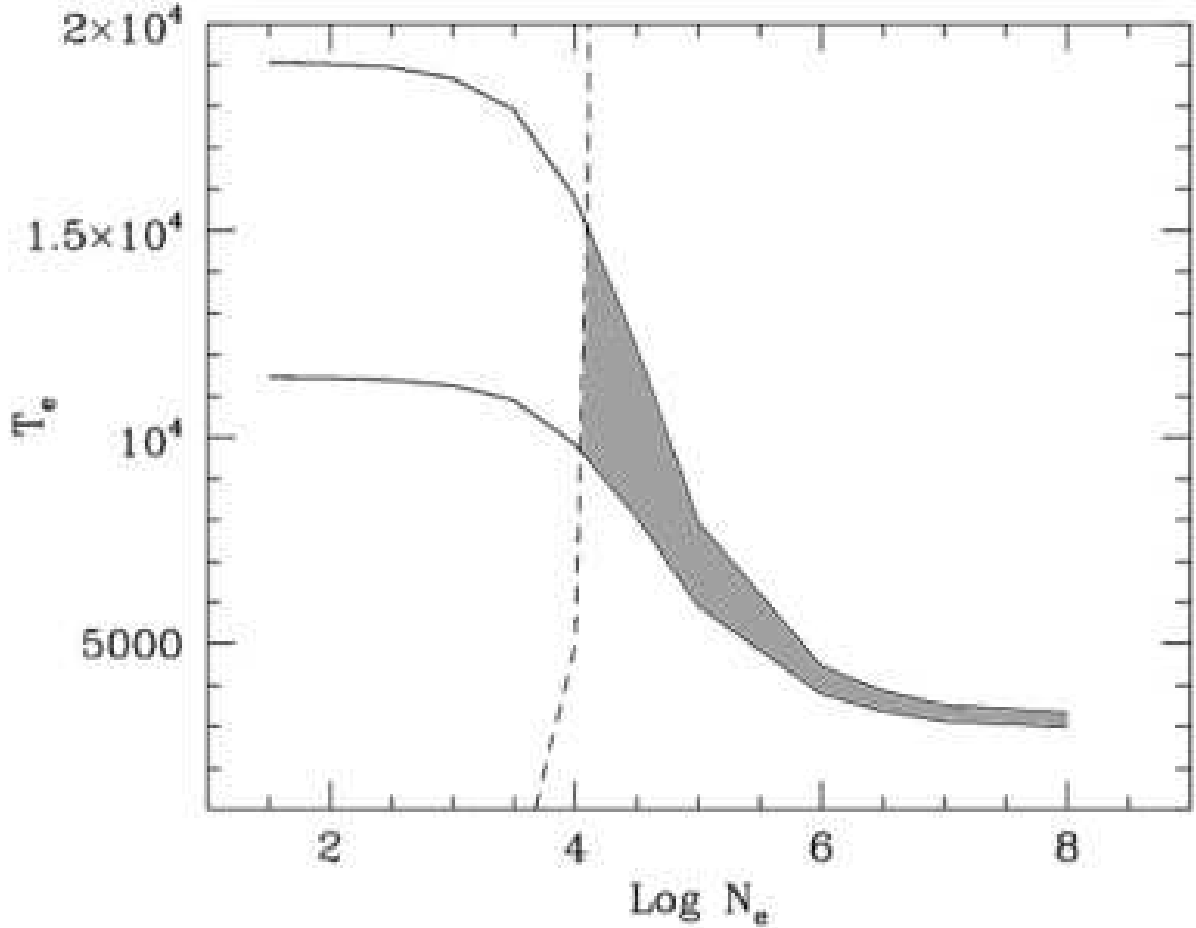} 
  \caption{{\elect} vs. {\elecd}. The continuous lines represent the theoretical curves of the {\elecd} and {\elect} pairs that reproduce the observed range of values 
of the [{\nii}] 5755/6584 line ratio measured in the proplyds. The dashed line represents the lower limit of the {\elecd} corresponding to the lowest value of the 
[{\sii}] 6731/6717 ratio measured in the proplyds. The grey area represents the permitted zone for the average spectral properties 
of the proplyds. The theoretical predictions have been constructed with emissivities calculated by the photoionization code PHOTO as described in 
\citet{stasinska05}, gently provided by Grazyna Stasi\'nska.}
  \label{col_dex}
\end{figure}

In order to further explore if the different behavior of the auroral and nebular {\elect}([{\nii}]) lines in some proplyds is due to 
collisional deexcitation, we have constructed Figure~\ref{col_dex}. In this figure, we show the 
theoretical curves of the {\elecd} and {\elect} pairs that reproduce the observed  
range of values of the [{\nii}] 5755/6584 line ratio in the proplyds, as well as the line of the lower limit of {\elecd} corresponding to the lowest value of the 
[{\sii}] 6731/6717 ratio measured in these objects. The theoretical predictions have been constructed with emissivities calculated by the photoionization  
code PHOTO as described in \citet{stasinska05}, gently provided by Grazyna Stasi\'nska. Unfortunately, 
the results of Figure~\ref{col_dex} are not conclusive, but considering that the estimated {\elecd} of the proplyds should be a lower limit of the true one, 
the permitted area of the diagram indicates that collisional deexcitation should be acting. Moreover, if that is correct, the true electron temperature should be lower 
than that actually indicated by {\elect}([{\nii}]), but perhaps not too different that the values corresponding to the surrounding ionized gas (never lower than 3000 K). 
It is important to note that the 
{\elect}([{\nii}]) values we obtain for the ``whole slit'' and the ``background gas'' spectra of positions 5 and 6 only show differences of the 
order of a few ten K. This indicates that the contribution of proplyds in the integrated spectrum does not produce a substantial increase of 
the electron temperature derived. On the other hand, in contrast with what happens in proplyds, the increase of {\elect}([{\nii}]) that we see in H-H objects 
does not seem to be related to collisional 
deexcitation. This is suggested because the H-H objects show lower values of {\elecd} and lower {\elect}([{\nii}]) peaks than the proplyds, 
their {\elect}([{\nii}]) and {\elect}([{\oiii}]) show 
similar enhancements (see figures 4b and 7b), and the spatial coincidence of conspicuous similar peaks in both [{\nii}] 5755 and 6584 \AA\ lines.  
Therefore, the [{\nii}] emission of the H-H objects should not be affected by substantial collisional deexcitation and their electron temperature is higher 
probably because the action of an additional source of heating, perhaps related to shock excitation. 

\subsection{Line Fluxes}
\label{line_fluxes} 

One of the main spectral properties of H-H objects is their strong emission in [{\feiii}] lines. In Figure 4c, we can see the spatial profile of 
$F$([{\feiii}] 4881)/$I$(H$\beta$) line ratio along slit position 3. In HH 204, 
the intensity of that line is a factor 5 brighter than in the Orion bar. That object also shows a similar enhancement in the [{\oi}] 6300 \AA\ line and more 
moderate ones in the [{\sii}] and [{\nii}] lines. In the case of HH 202 --observed in slit position 4--, the intensity of [{\feiii}] 4881 \AA\ line 
increases by a factor 10 (Figure 7c). The rest of the H-H observed: HH 204, HH 529, and HH 530 also show enhancenments in [{\feiii}], [{\oi}], [{\nii}], and 
[{\sii}] lines. These are common spectral features in shock-excited objects \citep[see][]{hartiganetal87}. In contrast, the proplyds show [{\oi}]\footnote{The [{\oi}] 
6300 \AA\ line could not be corrected from sky emission contribution. In any case, spatial variations of the intensity of this line along the slit can 
only be due to intrinsic nebular variations.} spikes of moderate 
intensity and reversed spikes in the [{\sii}] 6717, 6731 \AA\ lines, perhaps also due to collisional deexcitation 
because of the rather low critical densities of the upper levels of those [{\sii}] transitions. The [{\oi}] emission emerges from the photodissociation of OH in the 
H/H$_2$ front that lies close to the protoplanetary disk surface \citep{storzerhollenbach98}.

In figures 4d to 8d, we show the spatial profiles of the pure RLs of {\cii} 4267 \AA\ and {\oii} 4649 \AA\ along the slit positions. 
The spatial distribution of the {\cii} and {\oii} lines are very similar. However, in the cases of the slits positions 
passing through the center --specially in positions 1 and 6-- there is a slight decrease of the {\cii}/{\oii} ratio towards the central parts of the nebula. 
This variation could be due to the increase of the C$^{3+}$ ionization fraction near the Trapezium stars. 

In figures 4e to 8e, we show the spatial profiles of 
the {\oii} 4649 \AA\ and [{\oiii}] 4959 \AA\ lines, that show a fairly similar spatial distribution in all the slit positions. This behavior is very different to 
that observed in PNe 
\citep{liuetal00,garnettdinerstein01,krabbecopetti06} where the {\oii} line emission peaks closer to the central star than the [{\oiii}] line. 
We only find some localized enhancements of the [{\oiii}]/{\oii} ratio, 
which are related to the position of some proplyds (see figures 5e and 8e). These enhancements of the [{\oiii}]/{\oii} ratio are correlated with the increase of the continuum due to the 
emission of proplyds that show a large number of absorption features that produce a decrease in the intensity of the {\oii} 4649 \AA\ line (see \S\ref{adf}).  

\subsection{The Abundance Discrepancy Factor}
\label{adf} 

Finally, figures 4f to 8f show the spatial variation of the O$^{++}$/H$^+$ ratios obtained from CELs and RLs as well as the AD factor, 
(hereinafter ADF) defined as: 
\begin{eqnarray}
 {\rm ADF}({\rm O}^{++}) = {\rm log}({\rm O}^{++}/{\rm H}^+)_{\rm RLs} - {\rm log}({\rm O}^{++}/{\rm H}^+)_{\rm CELs} ,
\end{eqnarray}
for all the slit positions. The most interesting result concerning these figures is that the ADF remains fairly constant 
along most of the zones of the nebula observed, showing values between 0.15 
and 0.20 dex, in agreement with the determinations by Esteban et al. (1998 and 2004) based on deep echelle spectrophotometry of selected small areas 
of the Orion nebula. It is necessary 
to note that the 
behavior of the ADF close to the proplyds is not confident because the {\oii} lines are not properly measured in most of these objects due to the 
aforementioned strong increase of the continuum that makes difficult the measurement of weak lines and the effect of possible absortion features of their 
spectra. In any case, Figure 5f shows a quite convincing decrease of about 0.10 dex in the O$^{++}$/H$^+$ ratios obtained from CELs at proplyd 177-341. 

Another remarkable feature of the ADF along the slits can also be seen in Figure 5f, where we find a   
slightly higher ADF in an area between 
proplyds 159-350 and 177-341. It is interesting to note that slit position 5 (Figure 8f) also shows the same higher values of the ADF between 
proplyd 159-350 and the area just at the north of 177-341. That common behavior in both slit positions indicates that the 
local increase of the ADF should be real. This enigmatic zone of relatively high ADF has an apparent diameter of about 30$''$ and is located about 23$''$ south of 
$\theta^1$ Ori C star. There is not an apparent morphological and/or kinematical feature related to this zone. 

\begin{figure}
  \includegraphics[angle=0]{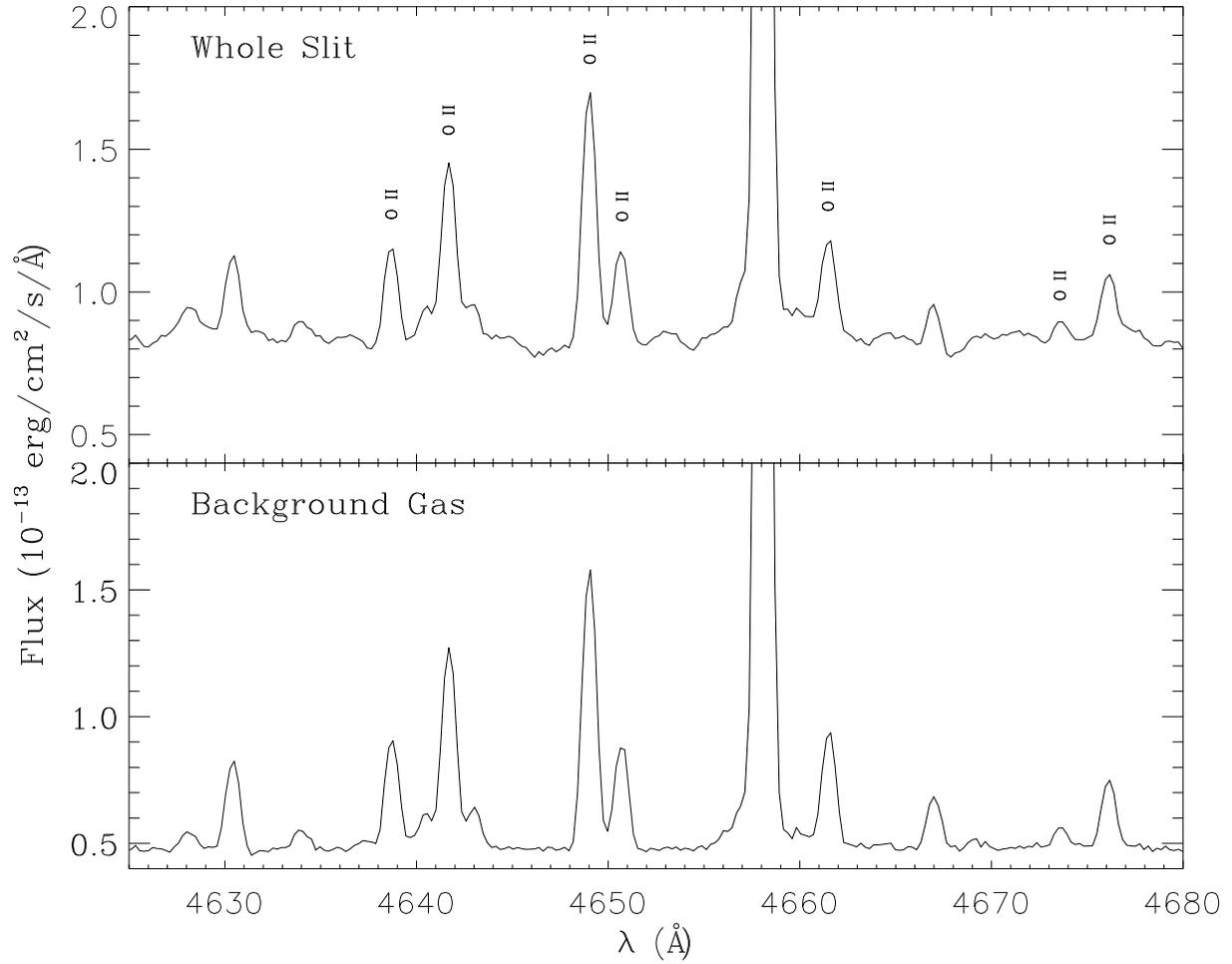} 
  \caption{Comparison of the spectral area around the multiplet 1 of {\oii} for the ``whole slit" (up) and ``background gas" (down) spectra of slit position 5. 
Note the rather different contribution of the continuum emission. The ``whole slit" spectrum includes the emission of proplyds and the ``background gas" 
spectra do not.}
  \label{comp_oii}
\end{figure}

In Figures 5f to 8f we also include 
the values of the ADF corresponding to the ``whole slit" and the ``background gas" spectra. It can be seen that the ADF of the ``background gas"  
is consistent with the average value of the different individual apertures extracted from the slit positions. However, the ADF obtained for the ``whole slit" 
spectrum --which is almost coincident with that of the ``background gas" in slit position 1, 3, and 4-- is 
substantially lower --about 0.1 dex-- in slit positions 6 and 5 (figures 5f and 8f, respectively). 
The reason of this surprising result is most probably related to the strong contribution of the proplyds to the continuum of the integrated spectra and that this 
contribution is producing some absorption in the {\oii} lines. In fact, in slit positions 
5 and 6, the continuum adjacent to the {\oii} 4649 \AA\ line is about a factor 1.7 higher in the ``whole slit" spectra than in the ``background gas" one 
(see Figure~\ref{comp_oii} for slit position 5). This higher continuum is also associated with a decrease of the intensity ratios of {\oii} lines 
with respect to H$\beta$. For example, the   
$I$({\oii} 4649)/$I$(H$\beta$) ratio is about a 25\% lower in the 
``whole slit" than in the ``background gas" spectrum of slit position 6. 

In figures 4f and 7f, we can note  
that the ADF is particularly large --reaching even values up to 0.6 dex-- along HH 203,  HH 204 and HH 202. The large increase of the ADF is due to the low values of 
the O$^{++}$/H$^+$ ratios determined by CELs in these zones. In contrast, the O$^{++}$ abundances determined by RLs do not show so strong localized 
decrease. It is 
interesting to compare the behavior of the ADF at the H-H objects and at the Orion bar in Figure~\ref{sp_pos3}. In the bar, we can see a similar decrease 
of the O$^{++}$/H$^+$ ratios determined from both kinds of lines but producing an ADF similar to the mean value along the slit. As it has been discussed 
above, collisional deexcitation does not seem to affect the intensity of the nebular [{\nii}] lines at HH 203 and HH 204 and, therefore, considering the larger 
critical densities of the 
[{\oiii}] lines, this effect is even less likely to be producing the observed decrease of the O$^{++}$/H$^+$ ratios determined from CELs.  
The observed behavior of the ADF in the H-H objects could be explained because the presence of a localized heating due to a non-radiative process --most likely 
shock excitation-- that would lead to the derivation of a lower O$^{++}$/H$^+$ (CELs) ratio, and this seems to be the case for HH 204 that shows a conspicuous 
spike of {\elect}([{\oiii}]) (Figure 4b). However, this hypothesis fails to 
reproduce the presence of high ADFs in HH 203 and HH 202, where there are not spikes of {\elect}([{\oiii}]). Another explanation is the presence of a localized 
high $t^{\rm 2}$ in those particular zones but, unfortunately, we can not check this possibility with our data. Finally, the other H-H objects observed --HH 529 and 
HH 530-- do not show 
any distinguishable enhancement of the ADF. 
   
\section{Radial Distributions of Some Relevant Nebular Parameters}
\label{rad_dist}

\begin{figure*}
  \includegraphics[angle=90,scale=0.8]{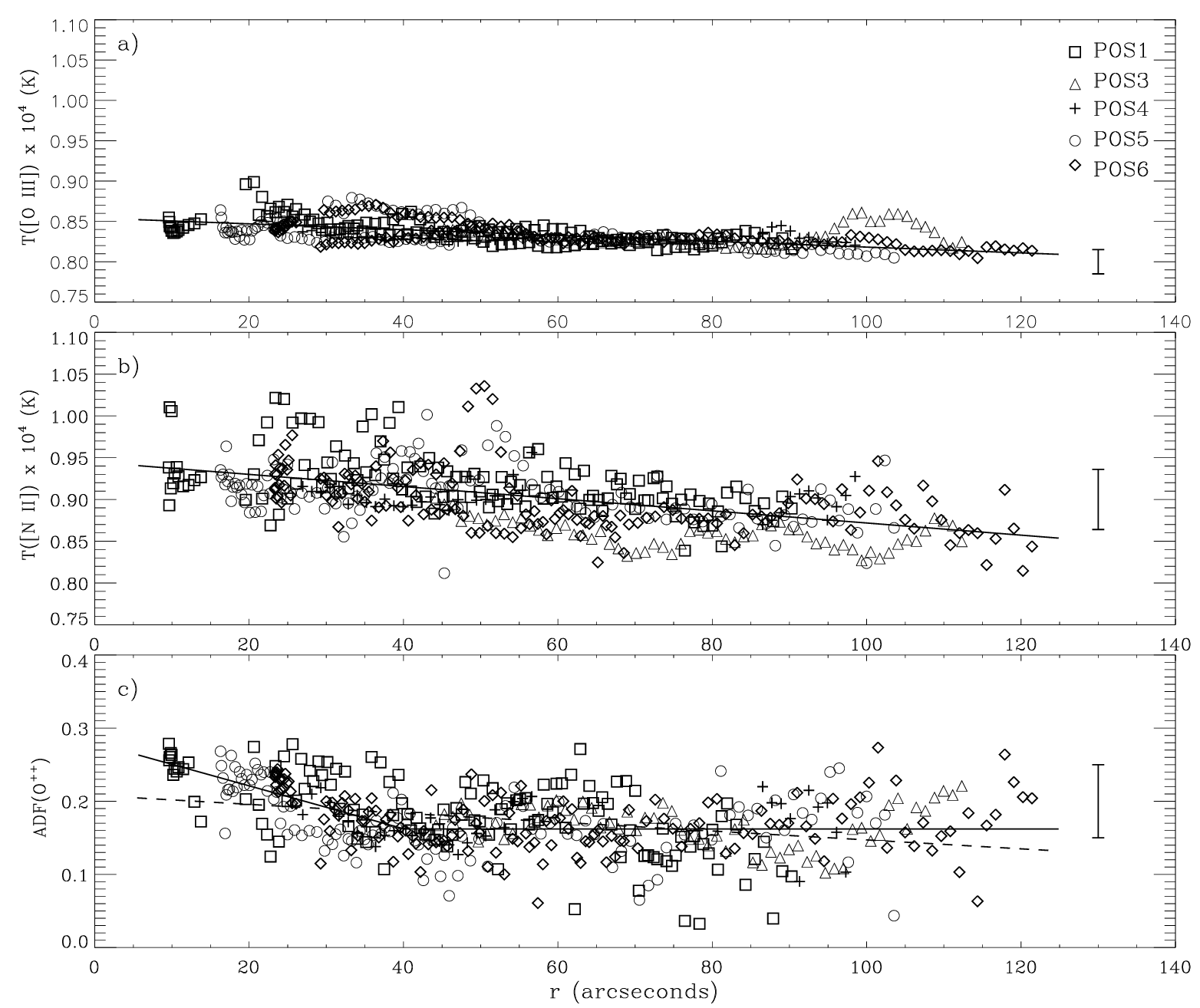} 
  \caption{Radial distribution of the values of several nebular parameters along the Orion nebula. We have included the data of the apertures 
of all slit positions excluding those associated with proplyds or Herbig-Haro objects, i.e. the figures show the behavior of the featureless 
ionized gas. The abscissa axis indicates the projected 
angular distance of the center of each individual 1D spectrum with respect to $\theta^1$ Ori C. Symbols are described in (a). 
Typical error bars are included in the different diagrams. (a) {\elect}([{\oiii}]), the continuous line represents a least-squared linear fit to the data. 
(b) {\elect}([{\nii}]), the continuous line represents 
a least-squared linear fit to the data. (c) ADF(O$^{++}$), the dashed line represents a least-squared linear fit to the data, the continuous line 
represents a divided least-squared linear fit to the data into two regions inside and outside 40$''$. The fits to the data are given in the text.}
  \label{rad_dist_fig}
\end{figure*}

We have constructed radial distribution diagrams of several nebular parameters combining the data of the different slit positions and projecting the 
angular distance of the center of each individual 1D spectrum with respect to $\theta^1$ Ori C (Figure~\ref{rad_dist_fig}). In these diagrams, we have only included the 
data of those apertures corresponding to the ``background gas", i.e. we have excluded all the points associated with proplyds or H-H objects. 
We include the same points in the three diagrams of Figure~\ref{rad_dist_fig} and only those for which the ADF was calculated. 
These diagrams  
permit to study possible large scale variations of the properties of the ionized gas across the nebula. 

In Figure 11a, we show the radial distribution of 
{\elect}([{\oiii}]), 
that shows a slight but clear general decrease with increasing distance from $\theta^1$ Ori C. In this figure, it can be seen a bump in 
the temperature distribution of slit position 3 between 90$''$ and 120$''$. This bump corresponds to the local increase of {\elect}([{\oiii}]) that 
occurs at the inner part of the Orion bar (see also Figure 4b). We have made a least-squared linear fit to the data shown in Figure 11a but excluding those 
points belonging to the Orion bar. The result is:
\begin{eqnarray}
 T_{\rm e}([{\rm O\medspace III}]) ({\rm K}) = 8540 - 3.6 \times r ,
\end{eqnarray}
where $r$ is the distance from $\theta^1$ Ori C in arcseconds. The uncertainty of the slope is 0.2 K arcsec$^{-1}$ and the Spearman's correlation coefficient is 
0.63 (hereinafter all the correlation coefficents we use are Spearman's ones). \citet{walteretal92} obtained a radial distribution of 
{\elect}([{\oiii}]) with a positive slope of 6.7 K arcsec$^{-1}$ in 
the northwestern quadrant of the nebula, fitting their own data and other from the literature. The result of these authors disagrees with the clear 
behavior shown by our data in Figure 11a, that seems to be independent 
of the location of each slit position and even includes zones of the northwestern quadrant --our slit position 4. On the other hand, 
\citet{odelletal03} do not find significant spatial variations of {\elect}([{\oiii}]) across the nebula in their high-resolution map of the {\foiii} ratio obtained with the 
WFPC at the $HST$. As we can see, the different results about the large-scale temperature variations in the Orion nebula are contradictory, however, 
we consider that our dataset is more reliable than the previous ones because it is based on homogeneous higher signal-to-noise ratio 
spectroscopic observations.  

In Figure 11b, it is evident that {\elect}([{\nii}]) also shows a radial decrease. The least-squared linear fit to the data gives: 
\begin{eqnarray}
 T_{\rm e}([{\rm N\medspace II}]) ({\rm K}) = 9460 - 7.8 \times r \ .
\end{eqnarray}

We can see that the slope of the fit is larger than that obtained for {\elect}([{\oiii}]) and, in this case, the result agrees qualitatively with 
the temperature gradient obtained by \citet{walteretal92} from the {\fnii} lines ($-$17.1 K arcsec$^{-1}$) in the inner 170$''$ of the nebula. 
The uncertainty of the slope of 
our fit is 0.6 K arcsec$^{-1}$ and its correlation coefficient is 0.52. \citet{sanchezetal07} obtain a bidimensional map of the spatial 
distribution of {\elect}([{\nii}]) finding also that this parameter is higher near the Trapezium stars and drops towards the outer zones of the nebula, 
in qualitative agreement with our result.  

Finally, in Figure 11c, we show 
the radial distribution of the ADF(O$^{++}$) across the nebula. In this figure, part of the relatively large dispersion is due to the large observation uncertainty in  
several apertures. In fact, most of the points with the highest and lowest values of the ADF correspond to low signal-to-noise ratio determinations at 
the edges of slit positions 1, 4, and 5. Despite the dispersion is 
relatively large, the first visual impression suggests that the ADF is rather constant 
or very slightly decreasing towards the outer parts of the nebula. In fact, the least-squares linear fit of all the points gives:
\begin{eqnarray}
 {\rm ADF(O^{++})} = 0.2050 - 0.0005 \times r \ ,
\end{eqnarray}
but with a very low correlation coefficient (0.32). However, other possibilities of fitting are possible. A more detailed inspection of Figure 11c 
shows that the ADF seems to increase toward the center in the innermost 40$''$ of the nebula and that this parameter becames basically constant for 
larger distances (the inner points of slit positions 1 and 5 consistently suggest the same tendency). We have divided the least-squared linear fit to 
the ADF into two regions, the first for the points in the inner 40$''$ and the second 
for the points outside this area --for which we have assumed a zero slope. The fits are the following:
\begin{eqnarray}
 {\rm ADF(O^{++})} = 0.279 - 0.003 \times r \ \ \  (r \leq 40'') \ ,
\end{eqnarray}
with a correlation coefficient of 0.58, and
\begin{eqnarray}
  {\rm ADF(O^{++})} = 0.163 \ \ \  (r > 40'') \ .
\end{eqnarray}
This apparent increase of the ADF toward the inner zones of the nebula can be ultimately related to two other tendencies we have found in our data. Firstly, 
the clear correlation illustrated in Figure 11a showing that {\elect}([{\oiii}]) is higher in the zones near the Trapezium stars and, secondly, the 
possible weak correlation between the ADF and {\elect}([{\oiii}]) that will be discussed in \S\ref{cor_ADF}. Therefore, the ADF seems to increase very slightly 
in the inner and systematically hotter zones of the nebula.  
If this behavior is real, it would indicate that whatever process is producing the ADF it increases somehow with the local ionization of the gas 
or that it depends on the distance between $\theta^1$ Ori C and the ionization front, from where most of the nebular 
emission comes. This last possibility arises because the blister geometry of the Orion nebula, the smaller distance of the Trapezium stars to the main ionization front 
occurs precisely behind that massive star cluster \citep[see][]{wenodell95}.

\section{Correlations Between the ADF and Other Nebular Properties}
\label{cor_ADF}

\begin{figure*}
  \includegraphics[angle=90,scale=0.8]{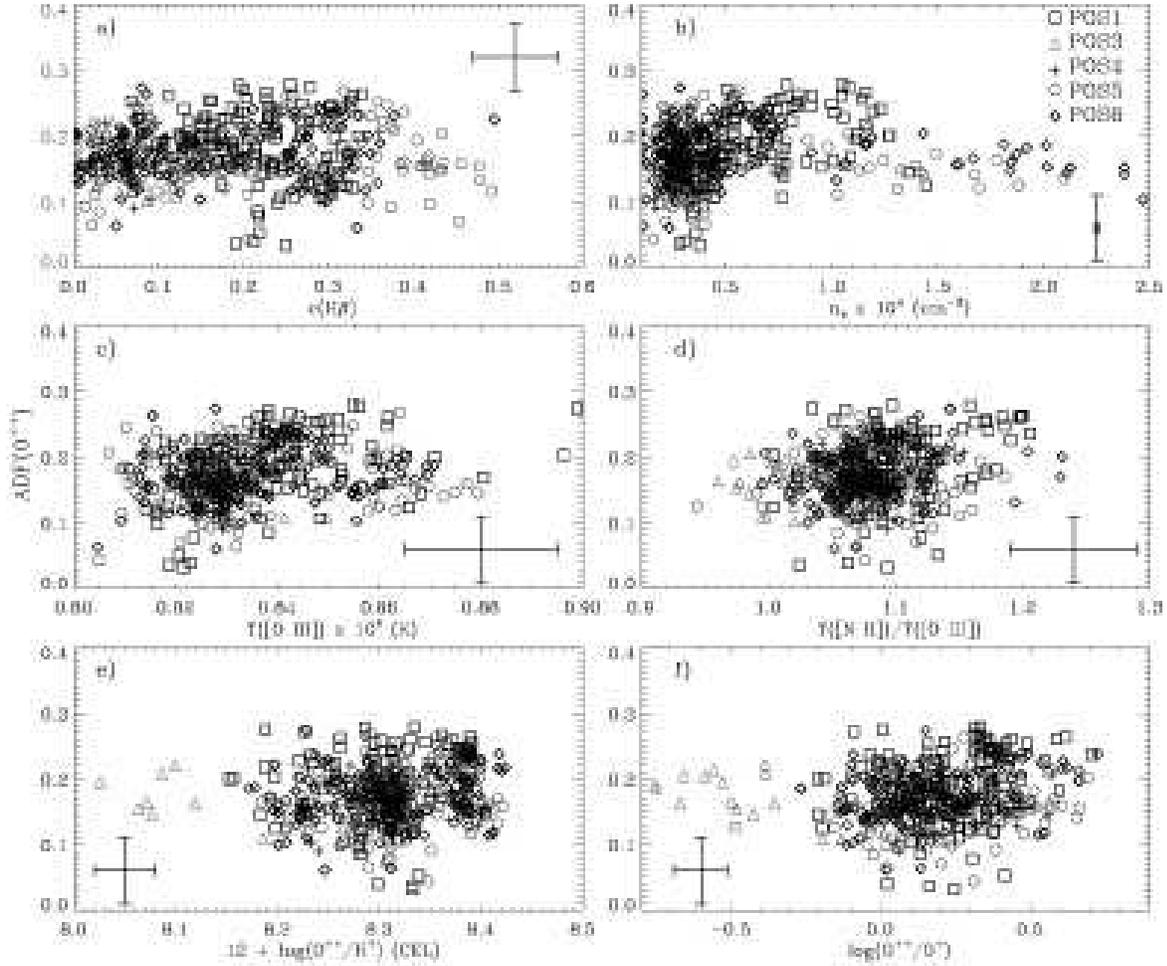} 
  \caption{The abundance discrepancy factor of O$^{++}$, ADF(O$^{++}$), vs. several nebular parameters. We include the data from all slit positions. 
Symbols are described in (b). Typical error bars are included in the different diagrams. (a) $c$(H$\beta$). (b) {\elecd}. (c) {\elect}([{\oiii}]). 
(d) {\elect}([{\nii}])/{\elect}([{\oiii}]) ratio. (e) O$^{++}$ abundance determined from CELs. (f) O$^{++}$/O$^{+}$ ratio.}
  \label{adf_correl}
\end{figure*}

In order to shed some light on the physical nature of the ADF problem, we have explored the relationship of this parameter with other nebular properties determined from 
our 1D spectra. Figure~\ref{adf_correl} illustrates the dependence of the ADF(O$^{++}$) with respect to a) $c$(H$\beta$), b) {\elecd}, c) {\elect}([{\oiii}]), 
d) {\elect}([{\nii}])/{\elect}([{\oiii}]) ratio, e) O$^{++}$ abundance determined from CELs, and f) O$^{++}$/O$^{+}$ ratio. As in Figure~\ref{rad_dist_fig}, 
we have only included 
the data of the apertures corresponding to the ``background gas". Figure 12a indicates that the ADF does not depend on the amount of dust present in the line of sight of 
each individual 1D spectrum. Also, there is no correlation between the ADF and {\elecd} as it shown in Figure 12b. However,  
from this figure it would seem that ADFs larger than about 
0.15 dex are only found in zones of {\elecd} $<$ 1.3 $\times$ 10$^4$ cm$^{-3}$. This apparent trend is incidentally produced by the apertures of slit position 1, 
which are located near $\theta^1$ Ori C and show the largest values of ADF, a behavior that does not seem to be related to {\elecd} but perhaps with the 
increase of {\elect}([{\oiii}]), as we will see below . 
The large dispersion of the ADF for the lowest densities 
shown in Figure 12b is an observational bias. The less dense zones show the lowest surface brightnesses and therefore the faintest spectra, implying 
larger uncertainties in their ADF determinations. The behavior of the ADF with respect to {\elect}([{\oiii}]) is 
shown in Figure 12c and seems to show an apparent weak positive correlation, although {\elect}([{\oiii}]) shows a rather narrow interval of values in the nebula. 

The other temperature indicator, {\elect}([{\nii}]), shows a similar behavior and this is reflected in Figure 12d, where we represent the ADF versus the 
{\elect}([{\nii}])/{\elect}([{\oiii}]) ratio. It is interesting to note 
that in Figure 12d the data 
points are distributed delineating the cloud of uncertainty around the average value of the ADF and {\elect}([{\nii}])/{\elect}([{\oiii}]),  
illustrating that the portion of the nebula observed shows a tight proportionality between both quantities. The lack of correlation between the ADF and the 
ratio of 
electron temperatures of ions located in different zones of the nebula indicates that the natural temperature gradients 
that should exist in ionized nebulae --due to the different spatial location of the main nebular coolants-- do not play a significant role in producing the 
observed ADF in the Orion nebula. This results is consistent with that found by \citet{garciarojasesteban07} in their analysis of integrated spectra of a sample of 
Galactic and 
extragalactic {\hii} regions. 
Figure 12e also shows a lack of correlation between the ADF and the O$^{++}$/H$^+$ ratio obtained from CELs --a similar diagram is found in the case of the 
O$^{++}$ abundance obtained from RLs, although the range of variation of the O$^{++}$/H$^+$ ratio is very narrow. Only the points belonging to the Orion bar show values 
of the O$^{++}$ abundance between 0.2 and 0.3 dex lower than the typical ones of the 
rest of the nebula, but their corresponding ADF is similar. Finally, Figure 12f represents the behavior of the ADF with respect to the O$^{++}$/O$^+$ ratio  
and it is evident that, despite the relatively large range of values covered by the ionic ratio, there is not an apparent trend between both quantities.

\section{Temperature Fluctuations}
\label{t2}

\begin{figure}
  \includegraphics[angle=0]{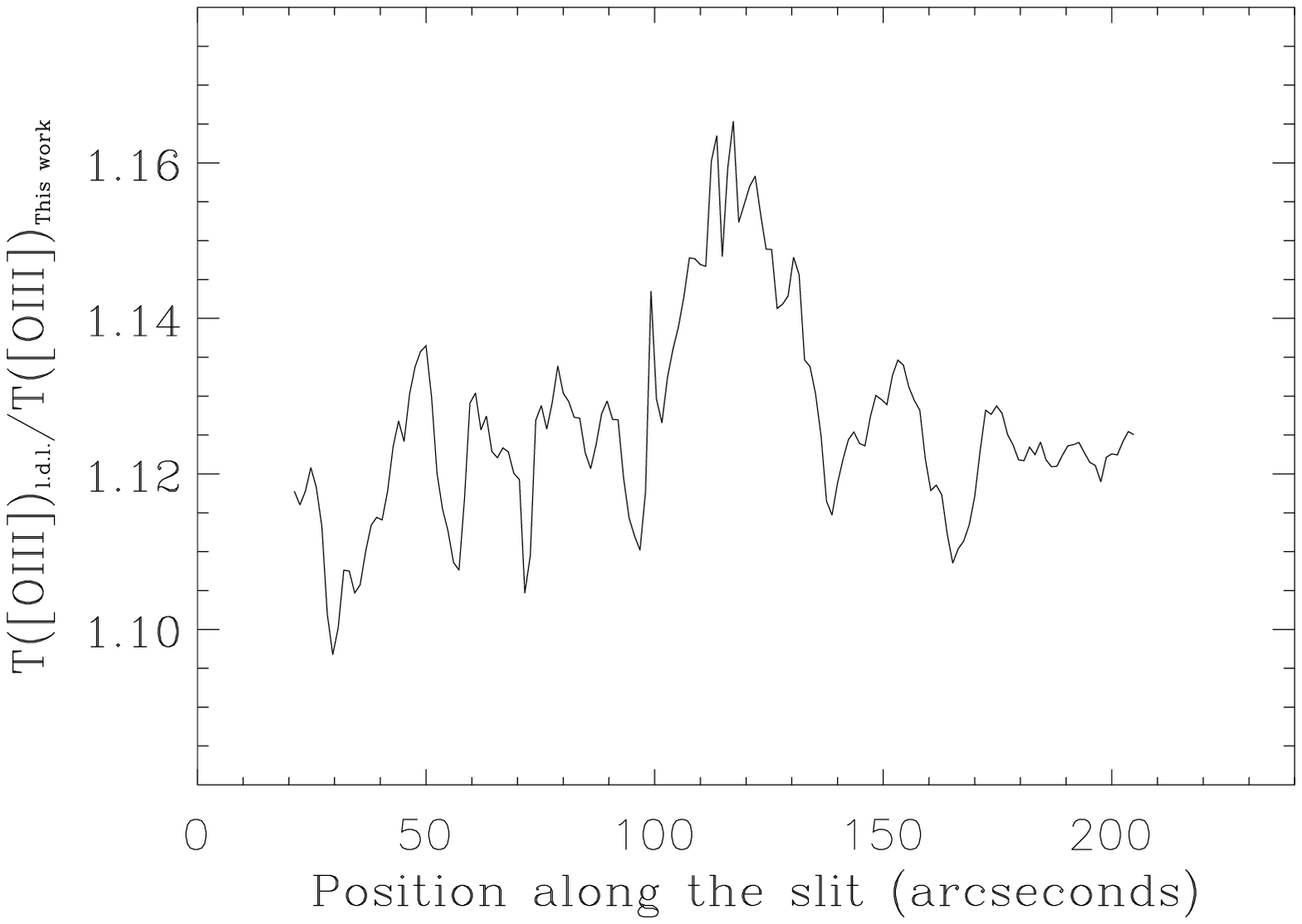} 
  \caption{Ratio of the {\elect}([{\oiii}]) computed using the low-density limit (l.d.l.) approximation equation used by \citet{rubinetal03} and our own 
determinations along slit position 6. Positional measurement along the slit goes
 from northeast to southwest (see Figure~\ref{slits}).}
  \label{temp_ratio}
\end{figure}

Following the formulation proposed by \citet{peimbert67}, the temperature fluctuation over the observed volume of a nebula can be parametrized in 
terms of the average temperature, $T_0$, and the mean-square electron temperature fluctuation, $t^{\rm 2}$, defined as:
\begin{eqnarray}
  T_{\rm 0} = \frac{\int{T_{\rm e}n_{\rm e}n_{\rm i}{\rm d}V}}{\int{n_{\rm e}n_{\rm i}{\rm d}V}} \ , 
\end{eqnarray}
and 
\begin{eqnarray}
  t^{\rm 2} = \frac{\int{(T_{\rm e}-T_{\rm 0})^2n_{\rm e}n_{\rm i}{\rm d}V}}{T_{\rm 0}^{\rm 2}\int{n_{\rm e}n_{\rm i}{\rm d}V}} \ , 
\end{eqnarray}
where $n_{\rm i}$ is the ion density. The integrations are calculated over the entire volume, and the element of volume, d$V$, can be expressed 
as d$l$d$A$, the product of the elements of length of column along the line of sight and surface area in the plane of the sky, respectively. 
Our spatially resolved spectroscopical data do not permit to obtain a direct determination of {\ts} along the line of sight. However, a discrete estimation 
of {\ts} in the plane of the sky --$t^{\rm 2}_A$-- can be obtained through the point-to-point determinations  
of the {\elect} we have obtained from the individual 1D spectra extracted along the slit positions. Following a similar procedure as 
\citet{liu98}, \citet{rubinetal03}, and \citet{krabbecopetti05} and assuming that {\elecd} $\approx$ $n$(H$^+$) in all the points of the nebula, we can 
compute $T_{{\rm 0},A}$ and $t^{\rm 2}_A$ using the following equations:
\begin{eqnarray}
  T_{{\rm 0},A} = \frac{\sum_{j}T_{{\rm e},j}n_{{\rm e},j}^2[n({\rm X}^{\rm +i})/n({\rm H}^+)]_j}{\sum_{j}n_{{\rm e},j}^2[n({\rm X}^{\rm +i})/n({\rm H}^+)]_j} \ ,
\end{eqnarray} 
and
\begin{eqnarray}
  t^{\rm 2}_A = \frac{\sum_{j}(T_{{\rm e},j}-T_{{\rm 0},A})^2n_{{\rm e},j}^2[n({\rm X}^{\rm +i})/n({\rm H}^+)]_j}{T_{{\rm 0},A}^{\rm 2}\sum_{j}n_{{\rm e},j}^2[n({\rm X}^{\rm +i})/n({\rm H}^+)]_j} \ , 
\end{eqnarray}
where $T_{{\rm e},j}$, $n_{{\rm e},j}$, and [$n({\rm X}^{\rm +i})/n({\rm H}^+)$]$_j$ are the electron temperature, electron density, and the 
ionic abundance of the X$^{+i}$ species in the   
$j$-th aperture extracted from a given slit position. The $T_{{\rm e},j}$ used in the equations above corresponds to average  
temperatures along the line of sight that crosses the nebula at each $j$-th aperture and it can be expressed as: 
\begin{eqnarray}
  T_{{\rm e},j} = \frac{\int{T_{\rm e}n_{\rm e}n_{\rm i}{\rm d}l}}{\int{n_{\rm e}n_{\rm i}{\rm d}l}} \ . 
\end{eqnarray}
Taking into account this fact, it is likely that the $t^{\rm 2}_A$ we obtain is substantially lower than {\ts} and, more strictly speaking, it should 
be considered a lower limit to {\ts} 
\citep[see further argumentation given by][]{rubinetal03,odelletal03}. 
In Table~\ref{ta2}, we 
summarize the values of $t^{\rm 2}_A$ we obtain for each slit position and for the two ions for which we have determinations of {\elect}. 
We have included all the points with {\elect} determination available in the sums, even those belonging to proplyds or H-H objects in order to 
explore the effect of these structures into the derivation of {\ts}. Part of the value of $t^{\rm 2}_A$ we calculate for each ion 
comes from errors in the measurement of the emission line ratios, so the intrinsic $t^{\rm 2}_A$ must be corrected by the relative mean quadratic error of 
the {\elect} measurements, $t^{\rm 2}_{Aer}$, with the simple relation $t^{\rm 2}_A$ $-$ $t^{\rm 2}_{Aer}$ \citep[see][]{odelletal03,krabbecopetti05}. 
The values of $t^{\rm 2}_{Aer}$ for each ion and slit position are also included in Table~\ref{ta2} and are always lower than the corresponding $t^{\rm 2}_A$ 
except in the case of slit position 3, where the errors are slightly higher, indicating that most of the temperature variation along slit position 3 
is produced by measurement uncertainties. 
The corrected $t^{\rm 2}_A$ we obtain is very low in all the cases, $t^{\rm 2}_A$(O$^{++}$) ranges from $\sim$0 to 0.0011 and $t^{\rm 2}_A$(N$^+$) 
from $\sim$0 to 0.0112. However, \citet{rubinetal03} obtain $t^{\rm 2}_A$(O$^{++}$) = 0.0068--0.0176 and $t^{\rm 2}_A$(N$^+$) = 0.0058--0.0175 for 
the different slit positions they observe in the Orion nebula, 
values which are relatively consistent with ours in the case of $t^{\rm 2}_A$(N$^+$) but considerably larger in the case of $t^{\rm 2}_A$(O$^{++}$). On the other hand, 
\citet{odelletal03} obtain $t^{\rm 2}_A$(O$^{++}$) = 0.0050--0.0156, values in good agreement with those obtained by \citet{rubinetal03} but also much higher  than our 
determinations. The reason of this discrepancy is difficult to ascertain. The resolution element of each set of observations is rather 
different, our 
apertures are of 1\farcs2 $\times$ 1\farcs03, and those of \citet{rubinetal03} and \citet{odelletal03} are 0\farcs5 $\times$ 0\farcs5 and 0\farcs1 $\times$ 0\farcs1, 
respectively. If the spatial resolution has something to do with the different $t^{\rm 2}_A$(O$^{++}$) obtained by us and the previous works, this 
would imply that the small-scale spatial variations of {\elect}([{\oiii}]) have a characteristic size between 0\farcs5 and 1\farcs0 and that those 
related to {\elect}([{\nii}]) are larger, at least of the order of the size of our apertures. In any case, this seems to us a rather unlikely scenario. Another difference 
between our determinations and those of \citet{rubinetal03} and \citet{odelletal03} is that those authors use a relation for determining {\elect} from 
the [{\oiii}] line 
ratio that is valid in the low-density limit, while our calculations are based on the solving of the statistical equilibrium equations for  
the {\elecd} measured for each particular aperture. We have explored the effect of this different procedure applying equation 5 of \citet{rubinetal03} 
--also used by \citet{odelletal03}-- to our data of slit position 6. In Figure~\ref{temp_ratio}, we show the ratio of the {\elect}([{\oiii}]) computed using the low-density limit 
approximation equation and our own 
determinations finding a systematical bias of about 1.12 and a small bump ($\sim$5\%) coincident with the zone of the highest density at the 
southwest of the Trapezium (see Figure 5a). The use of both different procedures do not substantially increase 
the {\elect} dispersion. In fact, using the low-density limit approximation, we obtain $t^{\rm 2}_A$(O$^{++}$) = 0.0008 for slit position 6, only 60\% larger than our 
calculations. Therefore, the use of a low-density 
limit approximation does not explain the large differences of $t^{\rm 2}_A$(O$^{++}$) find by us and the other authors. A last possibility is the different signal-to-noise ratio of the 
datasets we are comparing. As it was commented in \S~\ref{phys_cond}, this is a likely source of discrepancy in the case of the long-slit data of \citet{rubinetal03}, but 
this is more difficult to ascertain in the case of the data of \citet{odelletal03}. These authors estimate a representative 
probable error of about 4.2\% for their point-to-point {\elect}([{\oiii}]) determinations, which corresponds to a $t^{\rm 2}_{Aer}$ of about 0.0017, 
larger than our observational errors, but much lower than their nominal values of $t^{\rm 2}_A$(O$^{++}$), indicating that the temperature variations 
obtained by \citet{odelletal03} should be real if the errors are not largelly underestimated. Only observations combining very high signal-to-noise ratio and very high spatial 
resolution would ascertain this puzzle. 

\begin{deluxetable}{cccccc} 
\tabletypesize{\scriptsize}
\tablecaption{Values of $t^{\rm 2}_A$, $t^{\rm 2}_{Aer}$, and $t^{\rm 2}_l$
\label{ta2}}
\tablewidth{0pt}
\tablehead{
\colhead{Slit} & 
 & & & & 
\\
\colhead{Position} &   
\colhead{$t^{\rm 2}_A$(O$^{++}$)} &
\colhead{$t^{\rm 2}_{Aer}$(O$^{++}$)} &
\colhead{$t^{\rm 2}_A$(N$^+$)} &
\colhead{$t^{\rm 2}_{Aer}$(N$^+$)} &
\colhead{$<t^{\rm 2}_l$(O$^{++}$)$>$}} 
\startdata 
1 & 0.0013 & 0.0002 & 0.0127 & 0.0015 & 0.028 \\	
3 & 0.0004 & 0.0006 & 0.0010 & 0.0011 & 0.038 \\
4 & 0.0008 & 0.0003 & 0.0055 & 0.0016 & 0.029 \\
5 & 0.0004 & 0.0002 & 0.0126 & 0.0019 & 0.025 \\
6 & 0.0005 & 0.0002 & 0.0024 & 0.0017 & 0.025 
\enddata
\end{deluxetable}

As it has been stated before, our spatially-resolved spectroscopical data do not permit to determine {\ts} along the line of sight 
--that we will denote as $t^{\rm 2}_l$-- but an indirect estimate can be obtained assuming that the ADF is produced by the presence of 
such temperature variations. Although this is still a controversial possibility, there are some pieces of evidence indicating that this may 
be correct, at least in the case of {\hii} regions \citep[see][]{garciarojasesteban07}. 
\citet{odelletal03} show that the relation between $t^{\rm 2}_A$, $t^{\rm 2}_l$, and the total $t^{\rm 2}$ in three dimensions is:
\begin{eqnarray}
  t^{\rm 2} = t^{\rm 2}_A + <t^{\rm 2}_l>   \ , 
\end{eqnarray}
where $<$$t^{\rm 2}_l>$ is the average over all lines of sight. In Table~\ref{ta2}, we include the  
$<$$t^{\rm 2}_l$(O$^{++}$)$>$ we obtain for each slit position. These values have been estimated from the average ADF(O$^{++}$) of the individual 
apertures of each slit position. From the table, it can be seen that the $<$$t^{\rm 2}_l$(O$^{++}$)$>$ is rather similar 
for the different slit positions and consistent with previous 
determinations --between 0.018 and 0.028 \citep{estebanetal98,estebanetal04}-- except in the case of slit position 3, that shows a rather higher value. 
This large temperature fluctuation comes from the also larger 
ADFs we obtain at HH 203 and HH 204 (see Figure 4f). Taking into account that the values of $t$$^{\rm 2}_A$(O$^{++}$) are much lower than $<$$t^{\rm 2}_l$(O$^{++}$)$>$, it can be 
considered that 
{\ts}(O$^{++}$) $\approx$ $<$$t^{\rm 2}_l$(O$^{++}$)$>$. Following the argumentation of \citet{odelletal03}, this result would indicate that the hypothetical 
thermal inhomogeneities producing {\ts} should be small-scale ones and unresolved by our data, i.e. smaller than our spatial 
resolution limit of about 1$''$. On the other hand, our spatial resolution is unable to resolve the 
size of 10$^{13}$$-$10$^{15}$ cm that \citet{stasinskaetal07} derive for the metal-rich droplets they claim to be the most likely explanation of the AD. 

\section{Conclusions}
\label{conclu}
We have studied the spatial distribution of a large number of nebular quantities along five slit positions covering different morphological zones of the Orion nebula. 
The resolution element of the observations was 1\farcs2 $\times$ 1\farcs03. The studied quantities were c(H$\beta$), {\elecd}, {\elect}([{\nii}]), {\elect}([{\oiii}]), 
the intensity of several selected lines (H$\beta$, {\cii} 4267 \AA,
{\oii} 4649 \AA, [{\oiii}] 4959 \AA, [{\feiii}] 4881 \AA, [{\nii}] 5755 and 6584 \AA,
[{\oi}] 6300 \AA, and [{\sii}] 6717 + 6731 \AA), the O$^{++}$/H$^+$ ratio obtained from collisionally excited lines (CELs) and recombination lines (RLs), 
and the C$^{++}$/H$^+$ ratio obtained from RLs. The total number of apertures or 1D spectra extracted was 730. We have been able to determine the O$^{++}$/H$^+$ ratio 
from the faint RLs of this ion in a 92\% of the apertures. 

The spatial distribution of {\elecd} shows a large range of variation --larger than an order of mag\-ni\-tu\-de-- across the nebula, with local maxima associated with the 
position of protoplanetary disks (proplyds), Herbig-Haro objects, the Orion bar, and the brightest area of the nebula at the southwest of the Trapezium. The proplyds show 
quite prominent spikes of {\elect}([{\nii}]) and much lesser ones of {\elect}([{\oiii}]). This fact could be due to collisional deexcitation on the nebular lines 
of {\fnii} because of the high densities of these objects. Herbig-Haro objects also show somewhat higher values of {\elect}([{\nii}]) but, in this case, the origin 
could be related to extra heating of the gas due to shock excitation. The spatial distribution of the {\oii} 4949 \AA\ and {\foiii} 4959 \AA\ lines is fairly similar along  
all the slit positions, a very different behavior to that observed in planetary nebulae. We have found that the abundance discrepancy factor (ADF) of O$^{++}$ 
--the difference between the O$^{++}$ abundance determined from RLs and CELs-- remains, in general, rather constant along most of the observed areas of the nebula, 
showing values between 0.15 and 0.20 dex. However, there are some localized enhancements of the ADF, specially at the position of the Herbig-Haro objects HH 202, 
HH 203, and HH 204. 

The combined data of all slit positions indicate a clear decrease of {\elect}([{\nii}]) and {\elect}([{\oiii}]) with increasing distance from the main ionizing source of 
the nebula, $\theta^1$ Ori C. On the other hand, the radial distribution of the ADF shows a rather constant value across the nebula except at the inner 40$''$, 
where the ADF seems to increase very slightly toward $\theta^1$ Ori C. 

We have explored possible correlations between the ADF of O$^{++}$ and other nebular quantities, finding a possible very weak increase of the ADF for higher electron 
temperatures. There are not apparent trends between the ADF and c(H$\beta$), {\elecd}, the {\elect}([{\nii}])/{\elect}([{\oiii}]) ratio, O$^{++}$ abundance, and the 
O$^{++}$/O$^+$ ratio. 

Our spatially resolved spectroscopy allows to estimate the value of the mean-square electron temperature fluctuation in the plane of the sky, a lower 
limit to the traditional {\ts} parameter. We find very low values in all cases, result that is in contradiction with previous estimates from the literature. 
Our results indicate that the hypothetical thermal inhomogeneities --if they exist-- should be lower than our spatial resolution limit of about 1$''$. 

It is clear that further studies on the  {\elect}, chemical abundances, and ADF distributions at sub-arcsec spatial scales are necessary in trying to disentangle (a) whether 
small spatial scale temperature fluctuations and/or metal-rich droplets are really present in the Orion nebula and HII regions in general, and (b)  
the origin of AD problem and its possible relation with {\ts} and other nebular properties. The observations needed for this task are very difficult even for ground-based 
large-aperture telescopes and, by now, unfeasible with the current space telescopes and their available instrumentation. 

We thank G. Stasi\'nska and M. Rodr\'{\i}guez for their fruitful comments and help. We are grateful to the referee, Y. Tsamis, for his careful reading of the paper and his comments. This work has been funded by the Spanish Ministerio de Ciencia y Tecnología (MCyT) under project AYA2004-07466.

\end{document}